\documentclass{emulateapj}

\usepackage{epsfig}
\usepackage{amsmath}

\begin{document}

\newcommand{\DN}{${\rm D_N}$}
\newcommand{\Nseven}{${\rm N_{700}}$}
\newcommand{\kms}{km s$^{-1}$}
\newcommand{\kpc}{h$^{-1}$ kpc}
\newcommand{\MgII}{Mg {\sc II}}
\newcommand{\ew}{W$_{\rm eq}$(2796)}
\newcommand{\ewMg}{W$_{\rm eq}$(MgI b)}
\newcommand{\ewHa}{W$_{\rm eq}$(H$\alpha$)}
\newcommand{\ewHd}{W$_{\rm eq}$(H$\delta$)}
\newcommand{\ewNa}{W$_{\rm eq}$(NaD)}
\newcommand{\mew}{\rm W_{eq}(2796)}
\newcommand{\mewMg}{\rm W_{eq}(MgI b)}
\newcommand{\mewNa}{\rm W_{eq}(NaD)}
\newcommand{\Mrh}{M$_{\rm r} + 5 \log{\rm h}$}
\newcommand{\NaD}{Na {\sc I} D}
\newcommand{\MgIb}{Mg {\sc I} b}

\shorttitle{Mg {\sc II} Absorption at z $\sim$ 0.1}
\shortauthors{Barton et al.}

\title{Mg {\sc II} Absorption Characteristics of a Volume-Limited
Sample of Galaxies at z $\sim$ 0.1\altaffilmark{1}}

\author{\sc Elizabeth J. Barton\altaffilmark{2}
and Jeff Cooke\altaffilmark{2,3}}

\altaffiltext{1}{The data presented herein
were obtained at the W.M. Keck Observatory, which is operated as a
scientific partnership among the California Institute of Technology,
the University of California and the National Aeronautics and Space
Administration. The Observatory was made possible by the generous
financial support of the W.M. Keck Foundation.}

\altaffiltext{2}{Center for Cosmology, Department of Physics and
Astronomy, University of California, Irvine,  CA 92697-4575; ebarton@uci.edu, 
cooke@uci.edu}

\altaffiltext{3}{Gary McCue Postdoctoral Fellow}

\begin{abstract}

We present an initial survey of Mg {\sc II} absorption characteristics
in the halos of a carefully constructed, volume-limited subsample of
galaxies embedded in the spectroscopic part of the Sloan Digital Sky
Survey.  We observed quasars near sightlines to 20 low-redshift ($z
\sim 0.1$), luminous (\Mrh~$\leq -20.5$) galaxies in SDSS DR4 and DR6
with the LRIS-B spectrograph on the Keck I telescope.  The primary
systematic criteria for the targeted galaxies are a redshift $z
\gtrsim 0.1$ and the presence of an appropriate bright background
quasar within a projected 75 h$^{-1}$ kpc of its center, although
we preferentially sample galaxies with lower impact parameters
and slightly more star formation within this range.  Of the observed
systems, six exhibit strong [$\mew \geq 0.3$~\AA] Mg {\sc II}
absorption at the galaxy's redshift, six systems have upper limits
which preclude strong \MgII\ absorption, while the remaining
observations rule out very strong [$\mew \geq 1-2$~\AA] absorption.
The absorbers fall at higher impact parameters than many non-absorber
sightlines, indicating a covering fraction $f_c \lesssim 0.4$ for
$\geq 0.3$~\AA\ absorbers at $z \sim 0.1$, even at impact parameters
$\leq$35 h$^{-1}$ kpc (f$_{\rm c} \sim 0.25$).  The data are
consistent with a possible dependence of covering fraction and/or
absorption halo size on the environment or star-forming properties of
the central galaxy.

\end{abstract}


\section{Introduction} \label{sec:intro}

Because gas consumption timescales in galaxies are short, massive
galaxies must have large reservoirs of gas in order to fuel star
formation throughout cosmic time.  These reservoirs are likely quite
diffuse and extended, and are therefore difficult to probe directly.
Absorption lines in background quasars near galaxy sightlines allow a
unique probe of this gas \citep[e.g.,][and many
more]{Bergeron91,Steidel92,Steidel94,Steidel95,Chen01}.

Absorption-line studies in transitions like \MgII\ have revealed a
wealth of information about the absorbing gas and its relationship to
galaxies \citep[e.g.,][]{Bergeron91, Steidel94, Churchill96,
Nestor05,Zibetti05, Nestor06,Nestor07,Zibetti07}.  These results
suggest that many --- if not all --- luminous
intermediate-redshift galaxies are surrounded by \MgII-enriched gas
that extends out to at least $\sim$40 h$^{-1}$ kpc for luminous
systems \citep[$\sim$ 42 h$^{-1}$ kpc if $\Omega_{\Lambda} =
0.7$;][]{Steidel95} and likely much further in some cases
\citep{Churchill05, Kacprzak08}.  Estimates of the covering fractions
of these gaseous halos range from $0.17 < f_c \leq 1$
\citep{Steidel95, Bechtold92, Bowen95, Kacprzak08} in ``strong'' [\ew\
$> 0.3 \AA$] absorption systems.  This threshold for strong absorbers
is common and physically based; strong systems differ in character
from weaker systems, which have higher covering fractions and
different kinematics \citep{Churchill99,Churchill05, Nestor05,
Nestor06}.

If gaseous halos around galaxies are reservoirs for star formation
\citep{Maller04}, the observed properties of galaxies likely depend on
whether their gas halos are intact.  Processes that may inhibit star
formation in satellite galaxies include ``strangulation,'' where the
outer hot gas halo of a satellite galaxy is removed by the surrounding
environment \citep[e.g.,][]{Larson80,Kawata08}. This removal prevents
the further infall of cool gas.  A more violent and immediate process
is ram pressure stripping, which acts to remove cold, star-forming gas
from galaxies, possibly quenching star formation soon after infall
\citep[e.g.,][]{Gunn72,Moore96}.  These processes likely play a role
in the observed environmental dependence of galaxy properties like
color and morphology \citep[e.g.,][]{Gunn72,Dressler80,Postman84,
Blanton05a}. If stripping of the outer gas reservoir does regulate
star formation, we might expect to find correlations between the
properties of the galaxies and those of their halos.  Many \MgII\
studies find dependence on galaxy luminosity or halo mass
\citep[e.g.,][]{Steidel94, Bouche06}.  Recently, \citet{Kacprzak07}
have shown that stronger absorbers also preferentially reside in
asymmetric galaxies.  Many early studies did not reveal
trends with galaxy color or star-forming properties
\citep{Steidel94,Guillemin97}, although at least one study suggested
that trends are possible \citep{Bowen95}.

More recently, \citet{Zibetti07} study a sample of nearly 3000 \MgII\
systems with \ew\ $> 0.8$~\AA\ at $0.37 < z < 1$ by using
image-stacking techniques to probe the amount, spatial distribution,
and colors of the integrated galaxian light from the multiple, stacked
absorbers.  Their results demonstrate that \MgII\ absorption is color
dependent: the integrated galaxian light is bluer for stronger
absorbers and for galaxies that are closer to the quasar probes.
However, these results are aggregate and are luminosity-weighted; it
is difficult to disentangle the possible effects that result if, say,
\MgII\ absorption is more common in lower-luminosity blue galaxies.
Thus, while their creative statistical method has proven extremely
powerful, it cannot reveal a one-to-one relationship that may exist
between galaxy and \MgII\ properties.

\begin{figure}
\plotone{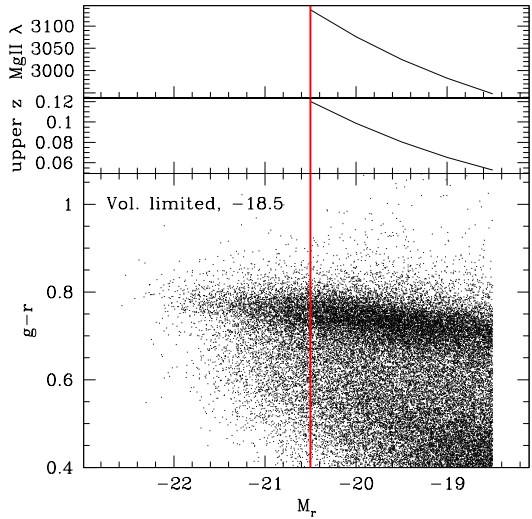}
\caption{Constructing a volume-limited sample of galaxies to probe for 
halo \MgII\ absorption. We illustrate the challenge inherent in the
use of SDSS to construct a sample for ground-based observations of 
halo\MgII\ absorption. ({\it Bottom}) We show the color-magnitude
distribution of a volume-limited sample to -18.5, showing both red and blue 
galaxies; observing as faint as possible is desirable
to sample the blue cloud.  As a function of absolute $r$-band
magnitude limit chosen from the SDSS, we show the upper redshift limit
({\it Middle}) and the expected \MgII\ absorption wavelength 
({\it Top}).  As we demonstrate in this study, wavelengths above
$\sim$3050 are largely accessible from the ground, suggesting that
volume-limited samples from SDSS with limits in the range of 
$\lesssim -21$ to $-20$ are possible to probe.  Here, we focus on 
the \Mrh $\lesssim -20.5$ sample.}
\label{fig:vol_lim}
\end{figure}

\begin{figure}
\plotone{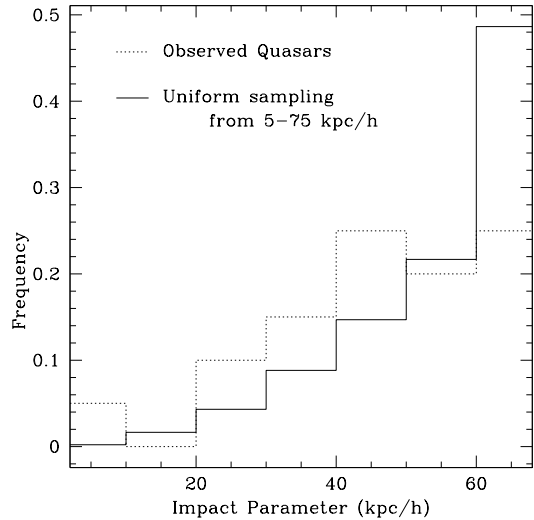}
\caption{Distribution of impact parameters probed in the 
study ({\it Dotted line}) compared to a uniform distribution from
5-75 h$^{-1}$ kpc.  The sample is deficient in galaxies with
impact parameters above $\sim$50h$^{-1}$ kpc (P$_{\rm K-S} = 0.01$).}
\label{fig:impact_histo}
\end{figure}

Most of our knowledge of \MgII\ absorbers comes from redshifts
$\gtrsim 0.4$, where the \MgII\ doublet ($\lambda$
2796.35,2803.53~\AA) is redshifted into the optical. Systems in
lower-redshift galaxies are difficult to observe from the ground and,
as a result, have not been studied as extensively.  To date,
detections indicate that the absorber abundances and covering
fractions decline somewhat.  Depending on their \ew, the decline is
not necessarily dramatic as redshifts approach zero
\citep{Bowen95,Nestor05, Nestor06}.

The vast majority of existing \MgII\ studies of individual
absorber-galaxy associations begin with quasar spectra and then use
photometric or spectroscopic redshifts to identify the nearby galaxies
responsible for the \MgII\ absorption \citep[e.g.,][and many
others]{Bergeron91,Steidel94, Churchill99}.  Studies that begin with
quasar spectra frequently focus on galaxies already known to have
\MgII\ absorption.  Some also tabulate galaxies that clearly lack
absorption, but \citet{Churchill05} note that focused ``control''
studies are biased toward quasars near non-absorbers, or absorbers
with patchier \MgII-absorbing gas. As \citet{Tripp05} point out,
``reversing'' the problem by identifying a controlled galaxy sample
first, and then finding a subset with background quasars, alleviates
significant selection effects.  \citet{Bowen95} present a Hubble Space
Telescope study of very nearby galaxies and \citet{Tripp05} present
preliminary results from a ``reverse'' study at $z\sim 0.2$ using
galaxies selected photometrically from the SDSS.

Here, we present a ``reverse'' study from a well-defined,
volume-limited spectroscopic sample of $\sim$L$^{\star}$ galaxies from
the Sloan Digital Sky Survey (SDSS).  Their basic properties, such as
environment, star-formation, and metallicity, can be placed into the
very well-defined context of the SDSS.  In \S~2 we describe our sample
selection and observational approach, and present the data.  \S~3
contains an analysis of our results, and we conclude in \S~4.

\section{Sample Selection and Data Reduction} \label{sec:data}

Our observations were designed to probe for the presence or absence of
\MgII\ absorption from the ground in a well-understood, volume-limited
{\it spectroscopic} survey of $\sim$L$^{\star}$ galaxies at $z \sim
0.11$, selected with the fewest biases possible.  The choice of
limiting absolute magnitude for the sample is informed by: (1) the
need to observe the strong [\ew\ $> 0.3$~\AA] \MgII\ lines as far
above the atmospheric cutoff as possible, and (2) the desire to
include lower luminosity galaxies, thus sampling both the red sequence
and the blue cloud.

We illustrate the relationships between \MgII\ wavelength, the upper
redshift limit, and the magnitude limit of a volume-limited sample
from SDSS in Fig.~\ref{fig:vol_lim}.  At $z \sim 0.1$, substantial
contributions to the blue cloud extend to roughly absolute magnitudes
of \Mrh~$\gtrsim -21$ \citep[e.g.,][]{Blanton03}.  A volume-limited
sample fainter than \Mrh$\sim -20$ has an upper redshift limit at
which \MgII\ is well below the atmospheric cutoff.  Thus, after
exploring the available quasars in the SDSS DR5 quasar catalog
\citep{Schneider07}, we focus on a limiting magnitude of M$_{\rm r} +
5 \log{h} \lesssim -20.5$, yielding a volume-limited sample with
enough quasar-galaxy pairs at $z \gtrsim 0.11$.  This study is
possible from the ground only with an optimally blue-sensitive
spectrograph like the Low-Resolution Imaging Spectrometer-Blue
channel \citep[LRIS-B][]{Oke95, McCarthy98}.  Few previous studies of
this nature have been conducted because the low spectroscopic
redshifts preclude ground-based observation of \MgII.

All but one of our targets originate from a volume-limited subsample
of the SDSS DR6 NYU Value-Added Galaxy Catalog (NYU-VAGC)
\citep{Adelman-McCarthy08, Blanton05b}.  The volume-limited galaxy
subsample constructed to M$_{\rm r} + 5 \log{h} \gtrsim -20.5$
includes 88,532 galaxies with redshifts $0.0044 \leq z \leq 0.116476.$
The colors and magnitudes we quote are k-corrected and Galactic
reddening corrected \citep{Blanton05b}; spectral line information
comes from the SpecLine data distributed via the SDSS web site.  We
ultimately select quasars at projected distances of $D \leq$ 75
h$^{-1}$ kpc in the SDSS DR5 quasar catalog.  The quasars satisfy m$_g
\leq 19.5$, with redshifts at which the observed-frame Lyman $\alpha$
emission from the quasar is blueward of the expected \MgII\ (2796~\AA)
absorption of the galaxy, allowing us to avoid contamination in the
quasar spectrum from intervening Lyman $\alpha$ forest systems.  We
observed a total of 20 quasar-galaxy pairings.  The signal-to-noise
ratios of the resulting observations vary depending on the seeing
conditions, the absorber redshift, and the flux of the quasar, as we
describe below.

\begin{table*}
{\scriptsize
\begin{center}
{\bf Table 1. Basic Properties of Targeted SDSS Galaxies}\\
\smallskip
\begin{tabular}{llrrrrrrl}
\hline
         & Galaxy Position & Galaxy & D              & QSO         & \multicolumn{2}{c}{Galaxy} & t$_{\rm exp}$ & Sample \\
QSO Name & (2000)          & z      & (kpc/h) & m$_{\rm g}$ & M$_{\rm R}$ & m$_{\rm r}$         & (sec.)        & Note  \\
\hline
SDSSJ030313.02-001457.4 & 03$^{\rm h}$ 03$^{\rm m}$ 13.27$^{\rm s}$ 00$^{\circ}$ 14$^{\prime}$ 20.5$^{\prime\prime}$ & 0.1049 & 50 & 18.40 & -21.38 & 16.38 & 3600 & DR4 only \\ 
SDSSJ080409.23+385348.8 & 08 04 11.10  38 53 17.8 & 0.0979 & 48 & 17.84 & -21.21 & 16.36 & 1800 &  \\ 
SDSSJ080814.69+475244.6 & 08 08 13.79  47 51 53.8 & 0.1094 & 71 & 17.89 & -20.55 & 17.31 & 1800 &  \\ 
SDSSJ081420.19+383408.3 & 08 14 21.97  38 33 49.2 & 0.0978 & 35 & 18.06 & -20.89 & 16.67 & 3600 &  \\ 
SDSSJ081940.82+443649.6 & 08 19 38.29  44 36 27.0 & 0.1116 & 49 & 19.05 & -20.68 & 17.27 & 2700 &  \\ 
SDSSJ090558.34+504538.0 & 09 05 55.34  50 45 16.4 & 0.1062 & 48 & 18.04 & -21.06 & 16.70 & 1800 &  \\ 
SDSSJ091119.16+031152.9 & 09 11 16.76  03 12 10.7 & 0.0963 & 50 & 17.43 & -20.79 & 16.75 & 3600 &  \\ 
SDSSJ092300.67+075108.2 & 09 23 01.04  07 51 05.1 & 0.1039 &  8 & 18.51 & -21.33 & 16.43 & 3600 &  \\ 
SDSSJ102216.13+621836.7 & 10 22 17.72  62 19 07.8 & 0.1162 & 48 & 19.34 & -20.66 & 17.28 & 2400 &  \\ 
SDSSJ102751.62+104532.6 & 10 27 49.99  10 46 05.2 & 0.1093 & 56 & 18.58 & -21.28 & 16.58 & 3600 &  \\ 
SDSSJ102847.00+391800.5 & 10 28 46.44  39 18 43.0 & 0.1135 & 62 & 17.09 & -20.87 & 17.06 & 1800 &  \\ 
SDSSJ104706.74+375315.4 & 10 47 06.24  37 52 25.8 & 0.1018 & 65 & 18.74 & -20.80 & 16.86 & 3300 &  \\ 
SDSSJ105033.08-001354.8 & 10 50 30.79  00 13 33.0 & 0.1155 & 59 & 18.81 & -21.35 & 16.63 & 3600 &  \\ 
SDSSJ111342.42-000730.7 & 11 13 42.52  00 07 55.6 & 0.1094 & 34 & 19.10 & -21.70 & 16.18 & 3600 &  \\ 
SDSSJ112613.52+352002.6 & 11 26 09.80  35 19 47.2 & 0.1117 & 68 & 17.79 & -20.88 & 17.03 & 3600 & DR6 only \\ 
SDSSJ114803.17+565411.4 & 11 48 03.81  56 54 25.6 & 0.1046 & 20 & 17.73 & -21.28 & 16.49 & 3000 &  \\ 
SDSSJ124914.11+392615.0 & 12 49 17.42  39 26 33.3 & 0.1133 & 60 & 18.69 & -20.63 & 17.27 & 3300 & DR6 only \\ 
SDSSJ140843.77+004730.4 & 14 08 42.24  00 47 35.0 & 0.1146 & 34 & 19.21 & -20.63 & 17.38 & 1800 &  \\ 
SDSSJ144033.82+044830.9 & 14 40 35.55  04 48 50.4 & 0.1129 & 46\tablenotemark{a} & 18.52 & -21.03 & 16.90 & 1800 &  \\ 
SDSSJ151541.23+334739.4 & 15 15 40.76  33 47 52.3 & 0.1156 & 20 & 18.57 & -20.59 & 17.39 & 1800 &  \\ 
\hline
\end{tabular}
\end{center}}
\caption{Properties of observations: (1) SDSS Name of quasar, (2) position of the targeted galaxy, (3)
redshift of targeted galaxy, (4) impact parameter, or projected on-sky distance between quasar
and center of targeted galaxy, (5) SDSS $g$-band magnitude of quasar, (6) and (7), absolute and
apparent $r$-band magnitudes of the targeted galaxy, (8) total exposure time on quasar, and (9) 
comments on sample membership for targeted galaxy.\label{tab:properties}}
\tablenotetext{a}{As noted in the text, there is a fainter companion galaxy (m$_{r} =18.1$) with a coincident redshift ($z =0.1128$) at 14$^{\rm h}$ 40$^{\rm m}$
34.6$^{\rm s}$ +04$^{\circ}$ 48$^{\prime}$ 25$^{\prime\prime}$, a projected distance of only $\sim$ 18.6 h$^{-1}$ kpc from the quasar. }
\end{table*}

\begin{table*}
{\footnotesize
\begin{center}
{\bf Table 2. Galaxy and Absorption Properties}\\
\smallskip
\begin{tabular}{lrrrrc}
\hline
         &        & M$_{\star}$            & EW(2796) & 3$\sigma$ Limit. & Point Style\\
QSO Name &  g-r   & (log M$_{\odot}$)      & (\AA)          & EW(2796) (\AA)  & in Figures  \\
\hline
\multicolumn{6}{c}{Detections}\\
\hline
SDSSJ091119.16+031152.9 &  0.692 & 10.47 & 0.50 & $0.23$ &  $\bullet$ \\ 
SDSSJ092300.67+075108.2 &  0.823 & 10.71 & 1.93 & $0.55$ &  $\bullet$ \\ 
SDSSJ102847.00+391800.5 &  0.450 & 10.15 & 0.36 & $0.09$ &  $\bullet$ \\ 
SDSSJ114803.17+565411.4 &  0.770 & 10.65 & 1.92 & $0.30$ &  $\bullet$ \\ 
SDSSJ144033.82+044830.9 &  0.603\footnotemark[a] & 10.42 & 1.24 & $0.15$ &  $\bullet$ \\ 
SDSSJ081420.19+383408.3 &  0.420 & 10.01 & 0.71 & $0.45$ &  $\bullet$ \\ 
\hline
\multicolumn{6}{c}{Potential Detections\tablenotemark{b}}\\
\hline
SDSSJ080409.23+385348.8 &  0.567 & 10.50 & 0.54 & $0.75$ & $\blacktriangle$ \\ 
SDSSJ081940.82+443649.6 &  0.722 & 10.39 & 0.40 & $0.27$ & $\blacktriangle$ \\ 
\hline
\multicolumn{6}{c}{Non-Detections of Strong Absorbers}\\
\hline
SDSSJ102751.62+104532.6 &  0.639 & 10.54 & $--$ & $< 0.23$ & $\circ$\\ 
SDSSJ105033.08-001354.8 &  0.611 & 10.53 & $--$ & $< 0.16$ & $\circ$\\ 
SDSSJ111342.42-000730.7 &  0.825 & 10.84 & $--$ & $< 0.25$ & $\circ$\\ 
SDSSJ112613.52+352002.6 &  0.749 & 10.50 & $--$ & $< 0.20$ & $\circ$\\ 
SDSSJ140843.77+004730.4 &  0.835 & 10.41 & $--$ & $< 0.27$ & $\circ$\\ 
SDSSJ151541.23+334739.4 &  0.678 & 10.38 & $--$ & $< 0.19$ & $\circ$ \\
\hline
\multicolumn{6}{c}{Weak Non-Detections}\\
\hline
SDSSJ030313.02-001457.4 &  0.744 & 10.68 & $--$ & $< 1.66$ & $\ast$ \\ 
SDSSJ080814.69+475244.6 &  0.715 & 10.35 & $--$ & $< 0.87$ &  $\ast$ \\ 
SDSSJ090558.34+504538.0 &  0.615 & 10.51 & $--$ & $< 0.55$ &  $\ast$ \\ 
SDSSJ102216.13+621836.7 &  0.435 & 10.09 & $--$ & $< 0.61$ &  $\ast$ \\ 
SDSSJ104706.74+375315.4 &  0.517 & 10.20 & $--$ & $< 1.36$ &  $\ast$ \\ 
SDSSJ124914.11+392615.0 &  0.452 & 10.12 & $--$ & $< 0.57$ &  $\ast$ \\ 
\hline
\end{tabular}
\end{center}}
\footnotetext[a]{The fainter companion galaxy
with a coincident redshift 18.6 h$^{-1}$ kpc from the quasar
has $g-r \sim 0.46$.}
\tablenotetext{b}{Significance formally $< 3\sigma$ in single
line but $\gtrsim 2.7\sigma$ in both lines; not readily apparent in
by-eye examination (see Fig.~\ref{fig:ambiguous}).\label{tab:absorbers}}
\caption{Description of results: (1) SDSS Name of quasar, (2) $g-$
color of the targeted galaxy, (3) stellar mass of the targeted galaxy,
(4) equivalent width of 2796~\AA\ \MgII\ line, (5) expected 2796~\AA\
upper limit for 3$\sigma$ detection, and (6) point style in Figures.}
\end{table*}

\begin{table*}
{\scriptsize
\begin{center}
{\bf Table 3.  Galaxy Spectral Properties from the SDSS}\\
\smallskip
\begin{tabular}{lrrrrrrr}
\hline
         & EW(H$\alpha$) &                           & EW(NaD)& EW(MgIb) & EW(H$\beta$) & EW(H$\delta$) & EW([OII]) \\
QSO Name & (\AA)         &  [NII]/H$\alpha$  & (\AA)  & (\AA)    & (\AA)        & (\AA)         & (\AA)     \\
\hline
SDSSJ030313.02-001457.4 &   1.19$\pm$0.20 &  $\gg 0.6$ & 3.46$\pm$0.17 & 7.47$\pm$0.34 &  2.35$\pm$0.21 & 1.58$\pm$0.32  & $--$ \\
SDSSJ080409.23+385348.8 & -21.39$\pm$0.28 & 0.34       & 3.10$\pm$0.20 & 3.62$\pm$0.39 & -2.68$\pm$0.18 & 2.94$\pm$0.42  &  -1.99$\pm$0.30 \\
SDSSJ080814.69+475244.6 &  -1.00$\pm$0.33 & 2.24       & 2.56$\pm$0.23 & 6.34$\pm$0.50 &  1.67$\pm$0.37 & 1.31$\pm$0.35  &  -6.52$\pm$0.55 \\
SDSSJ081420.19+383408.3 &  -9.67$\pm$0.23 & 0.45       & 1.29$\pm$0.26 & 3.65$\pm$0.51 & -0.97$\pm$0.15 & 3.62$\pm$0.40  &  -3.21$\pm$0.35 \\
SDSSJ081940.82+443649.6 &  -1.10$\pm$0.15 & 2.18       & 4.48$\pm$0.23 & 7.13$\pm$0.64 &  1.89$\pm$0.30 & 1.57$\pm$0.32  &  -3.27$\pm$0.51 \\
SDSSJ090558.34+504538.0 &  -1.86$\pm$0.33 & 1.21       & 2.99$\pm$0.39 & 10.43$\pm$1.50 &  3.04$\pm$0.90 & 3.38$\pm$1.78  &  -3.12$\pm$1.63 \\
SDSSJ091119.16+031152.9 &  -2.09$\pm$0.14 & 1.25       & 4.26$\pm$0.20 & 4.87$\pm$0.46 &  2.84$\pm$0.48 & 1.17$\pm$0.57  &  0.07$\pm$0.20 \\
SDSSJ092300.67+075108.2 &  -0.06$\pm$0.08 &  $\gg 0.6$ & 5.77$\pm$0.21 & 8.35$\pm$0.69 &  3.54$\pm$0.46 & 2.48$\pm$0.72  &  -2.68$\pm$0.88 \\
SDSSJ102216.13+621836.7 & -19.68$\pm$0.48 & 0.35       & 2.23$\pm$0.39 & 5.07$\pm$1.44 & -4.36$\pm$0.38 & 6.71$\pm$1.12  &  -7.59$\pm$0.79 \\
SDSSJ102751.62+104532.6 &  -1.74$\pm$0.28 & 1.80       & 2.35$\pm$0.38 & 7.40$\pm$0.70 &  2.64$\pm$0.57 & -0.07$\pm$0.25  &  -3.09$\pm$0.85 \\
SDSSJ102847.00+391800.5 & -31.35$\pm$0.49 & 0.35       & 1.86$\pm$0.37 & 1.80$\pm$0.38 & -5.25$\pm$0.27 & 0.17$\pm$0.13  &  -7.54$\pm$0.45 \\
SDSSJ104706.74+375315.4 & -24.59$\pm$0.50 & 0.76       & 0.33$\pm$0.20 & 3.89$\pm$0.73 & -13.04$\pm$0.95 & -0.24$\pm$0.22  &  -1.40$\pm$0.27 \\
SDSSJ105033.08-001354.8 & -6.32$\pm$0.23 & 0.55       & 3.38$\pm$0.18 & 5.92$\pm$0.43 &  0.88$\pm$0.15 & 2.10$\pm$0.40  &  -1.26$\pm$0.28 \\
SDSSJ111342.42-000730.7 &  0.14$\pm$0.05 & $\gg 0.6$  & 6.11$\pm$0.21 & 9.16$\pm$0.39 &  2.12$\pm$0.30 & 1.07$\pm$0.27  &  5.93$\pm$2.27 \\
SDSSJ112613.52+352002.6 & -1.31$\pm$1.14 & $--$       & 4.10$\pm$0.25 & 8.45$\pm$0.53 &  2.96$\pm$0.37 & 1.40$\pm$0.38  & 13.10$\pm$21.89 \\
SDSSJ114803.17+565411.4 & -0.85$\pm$0.23 & 3.44       & 6.71$\pm$0.29 & 8.47$\pm$0.74 & -0.16$\pm$0.13 & -0.21$\pm$0.39  &  -7.26$\pm$1.12 \\
SDSSJ124914.11+392615.0 & -53.00$\pm$5.00 & $--$       & $--$             & 1.87$\pm$0.37 & -12.04$\pm$0.30 &  -0.50$\pm$0.16  & -12.46$\pm$0.37 \\
SDSSJ140843.77+004730.4 & -0.36$\pm$0.15 &  0.6       & 4.63$\pm$0.35 & 7.38$\pm$0.79 & 2.25$\pm$0.44 & 0.49$\pm$0.40  & 1.89$\pm$1.37 \\
SDSSJ144033.82+044830.9 & -6.17$\pm$0.17 & 0.51       & 2.65$\pm$0.19 & 5.09$\pm$0.41 & -0.46$\pm$0.11 & 2.62$\pm$0.36  &  -3.02$\pm$0.45 \\
SDSSJ151541.23+334739.4 & -6.95$\pm$0.24 & 0.75       & 3.97$\pm$0.22 & 4.78$\pm$0.41 & 2.25$\pm$0.44 & 2.97$\pm$0.38  &  -2.29$\pm$0.25 \\
\hline
\end{tabular}
\caption{SDSS SpecLine properties of targeted galaxies: (1) SDSS Name of quasar, (2) H$\alpha$ equivalent
width, (3) ratio of [NII] 6586~\AA\ line to H$\alpha$, (4) - (8) equivalent widths of \NaD, \MgIb, H$\beta$,
H$\delta$, and [OII].\label{tab:spectra}}
\end{center}}
\end{table*}

\begin{figure}
\plotone{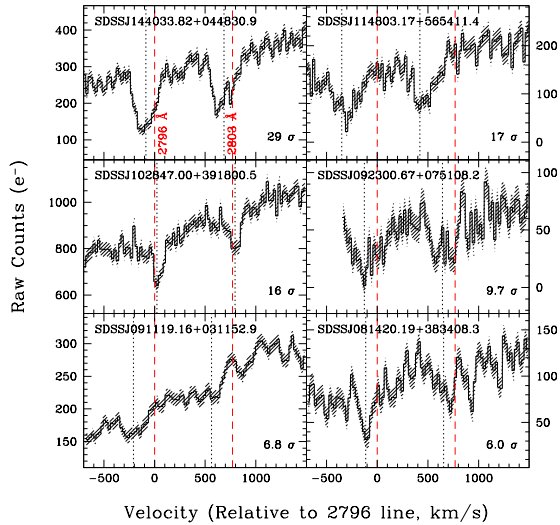}
\caption{The detections.  We plot raw object counts in detected
electrons as a function of effective rest-frame (peculiar) velocity
shift from the centroid of the 2796 \AA\ line. The shading indicates
the errors and vertical lines ({\it Dashed, red}) indicate the
expected positions of the MgII lines from the galaxies if they occur
at the redshift centroid; thinner ({\it dotted}) lines indicate the
velocity offset of the maximum detection signal.  We list the quasar
name in the upper left, and the significance of the detection in the
lower right. The bottom two spectra are box-car smoothed by 4 and 2
bins, respectively. }
\label{fig:detections}
\end{figure}

\begin{figure}
\plottwo{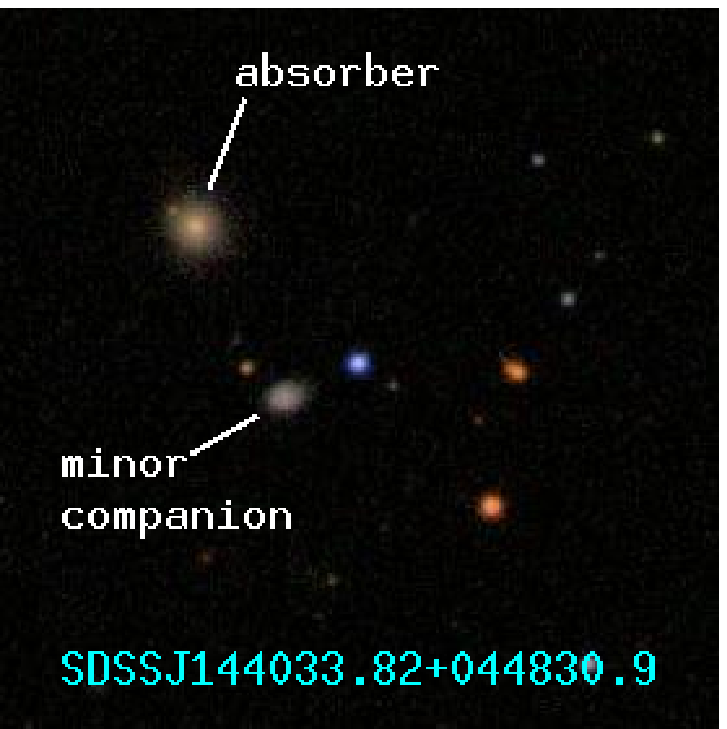}{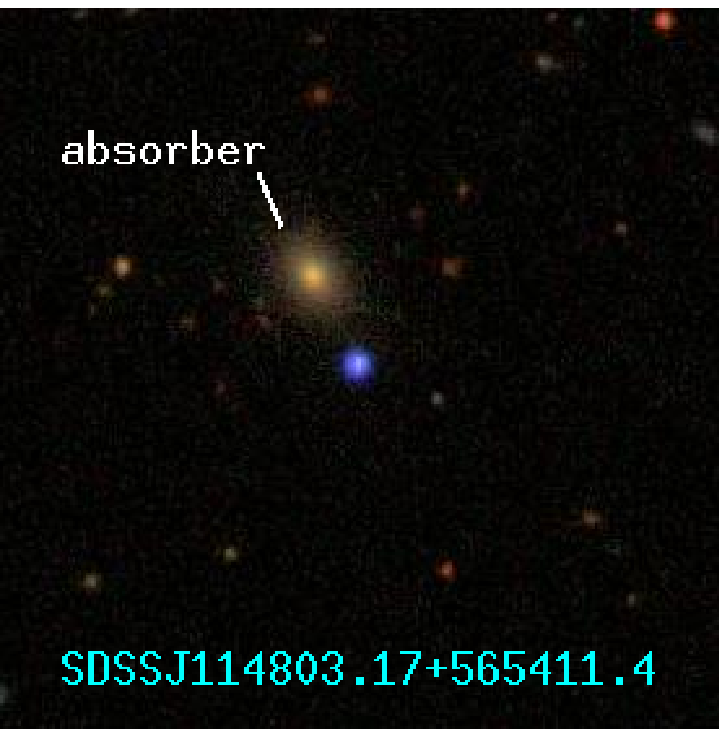}
\plottwo{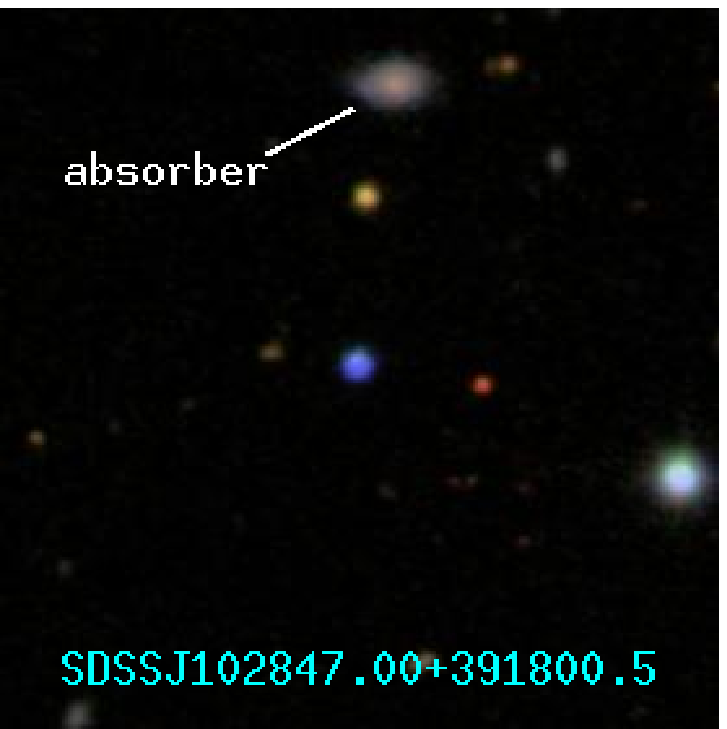}{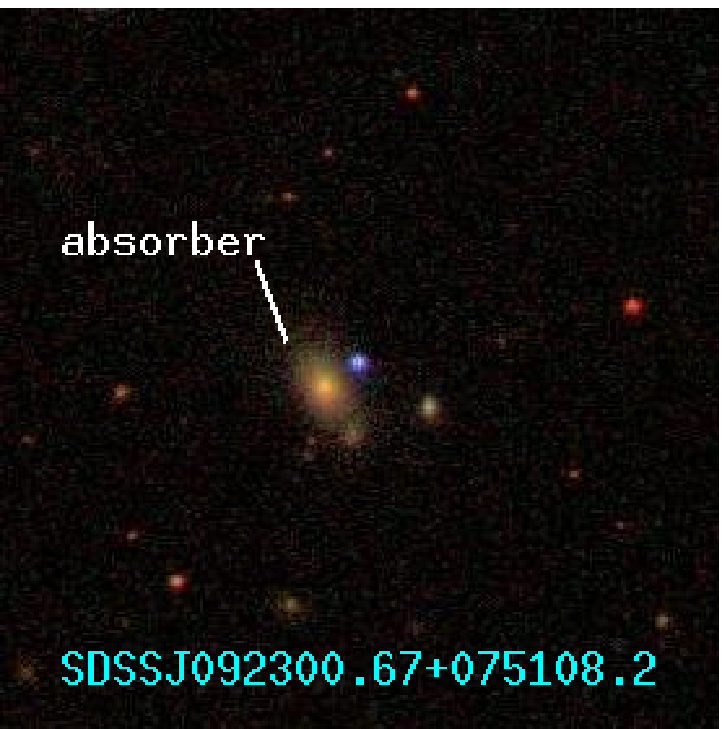}
\plottwo{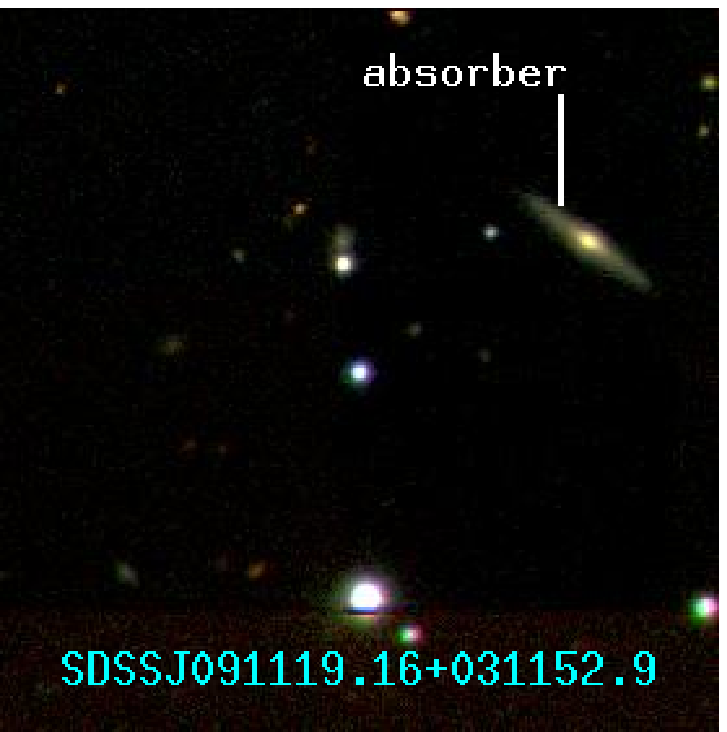}{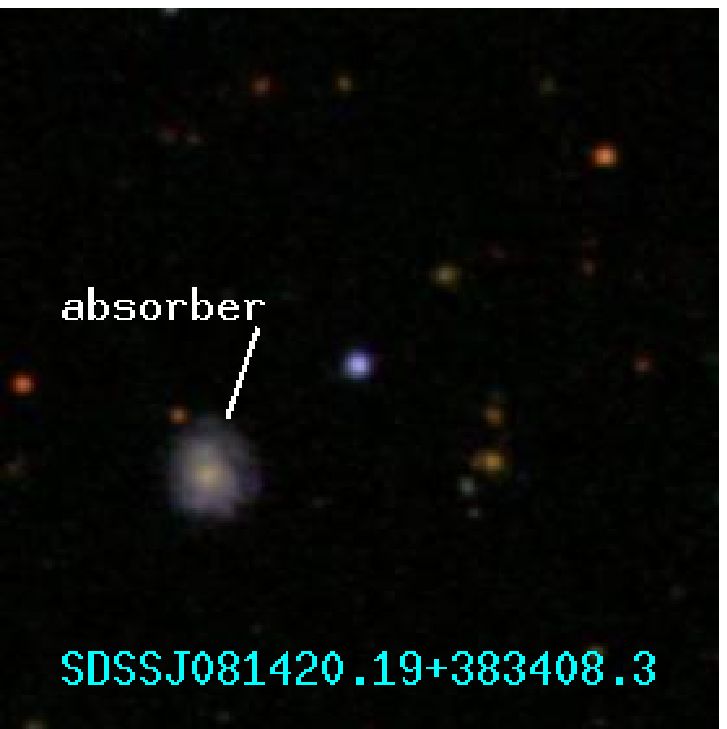}
\caption{The detected galaxies.  We show SDSS images of the quasars
and likely absorbers.  The central object is the quasar, and the 
images are 40$^{\prime\prime}$ on a side, corresponding ({\it Left to right}) 
to 58, 54, 58, 54, 50, and 51 h$^{-1}$ kpc, respectively.
}
\label{fig:detections_pix}
\end{figure}

\begin{figure}
\plotone{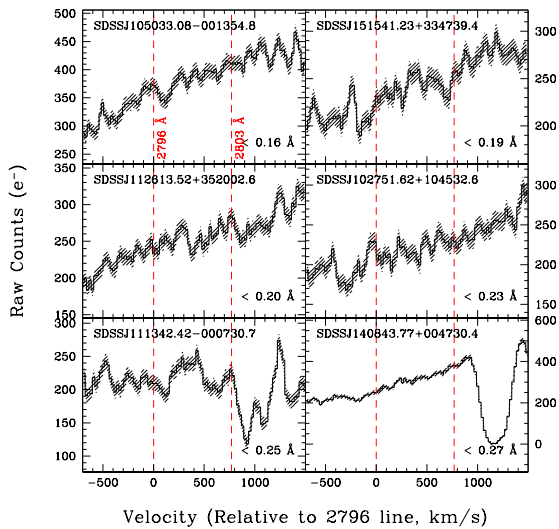}
\caption{The solid non-detections of $\geq 0.3$~\AA\ absorbers.  We
plot raw object counts in detected electrons as a function of
effective rest-frame (peculiar) velocity shift from the centroid of
the 2796 \AA\ line. The shading indicates the errors and vertical
lines ({\it Dashed, red}) indicate the expected positions of the MgII
lines from the galaxies if they occur at the redshift centroid. We
list the quasar name ({\it Upper left}), and
the 3$\sigma$ upper limit (rest-frame; {\it Lower right}). The spectra
are box-car smoothed by 3 bins. SDSSJ151541.23+334739.4 likely has an
absorber but it satisfies $\mew \leq 0.19$~\AA\ and is therefore not
a strong absorber. Note that if the feature at 100 \kms\ in
SDSSJ111342.42-000730.7 were the 2976~\AA\ line its equivalent width
would only be 0.2~\AA.}
\label{fig:non_detections}
\end{figure}

\begin{figure}[b]
\plottwo{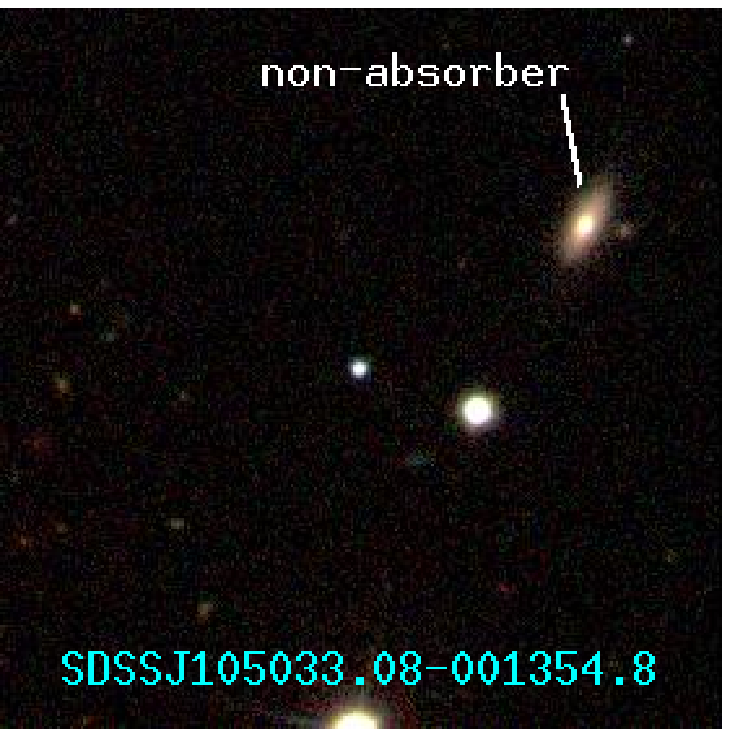}{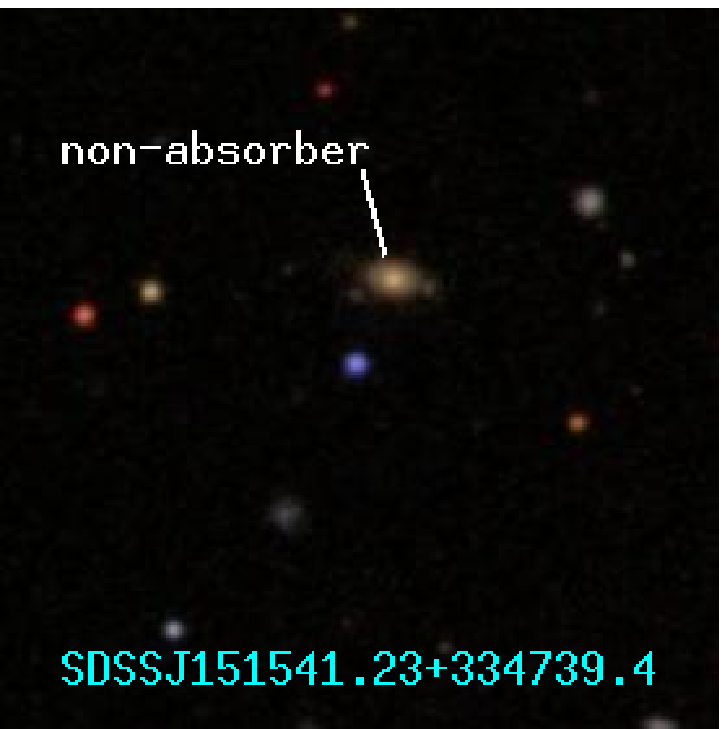}
\plottwo{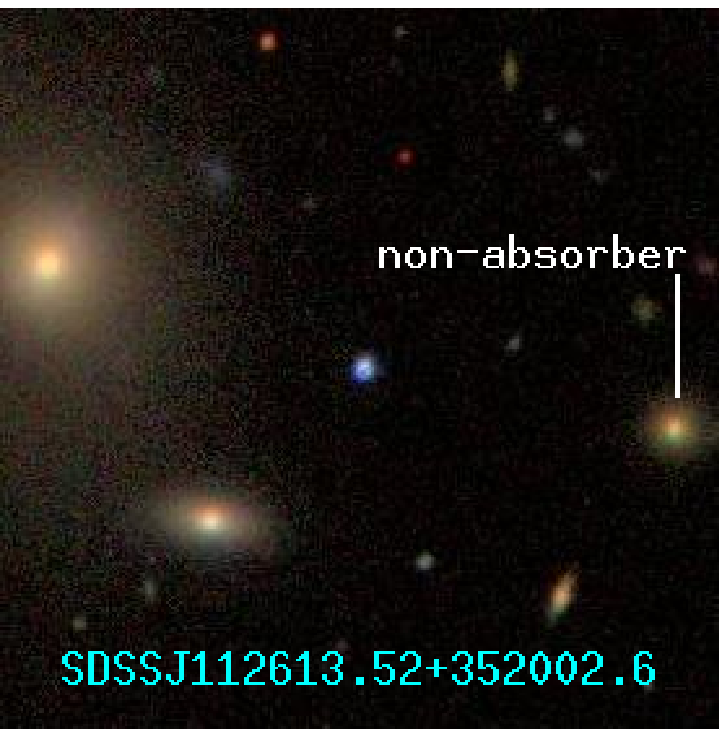}{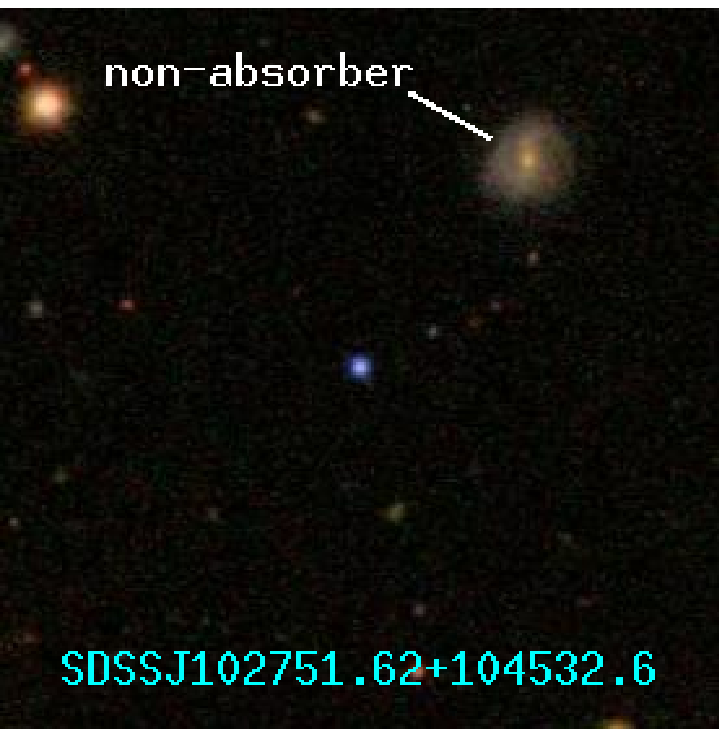}
\plottwo{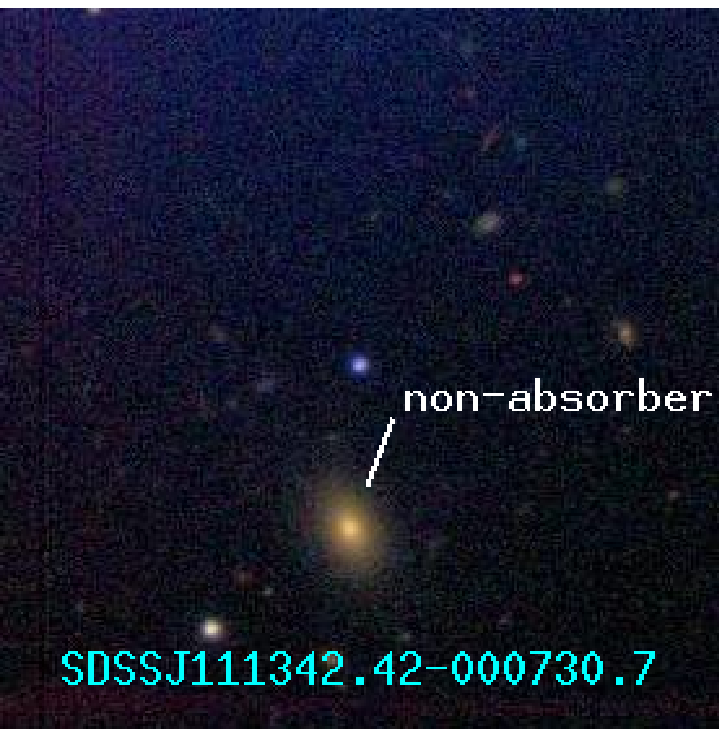}{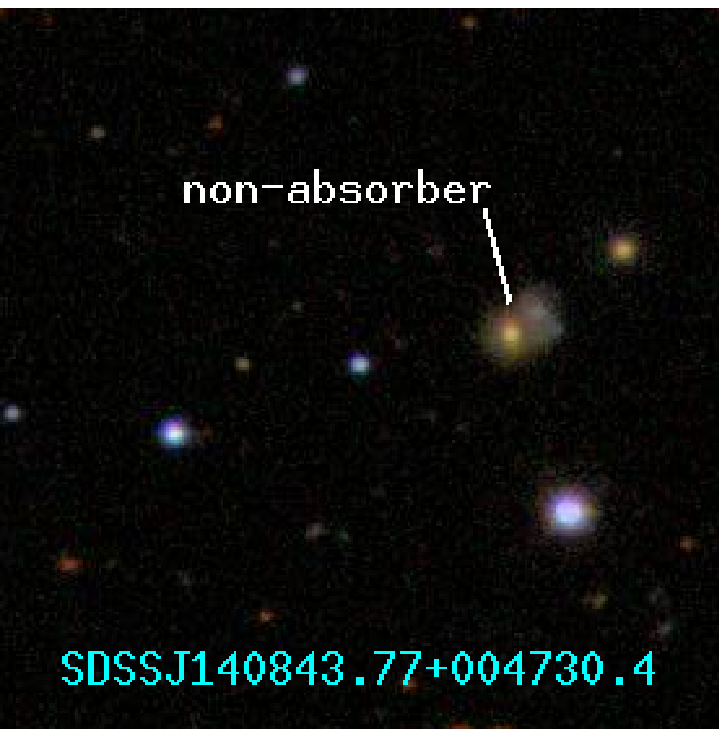}
\caption{The galaxies that were not detected.  We show SDSS
images of the quasars and galaxies with $\leq 0.3$~\AA\ upper limits
that are non-detections of strong absorbers.  The central object is
the quasar and the images are 40$^{\prime\prime}$ on a side,
corresponding ({\it Left to right}) to 59, 59, 57, 56, 56,, and
58 h$^{-1}$ kpc, respectively.}
\label{fig:nondetections_pix}
\end{figure}

Our observations began with a pilot study that sampled the SDSS VAGC
DR4 \citep{Adelman-McCarthy06}, with a magnitude limit that extended
to \Mrh$\leq -20$.  Because the lower-redshift systems were more
difficult to target than expected, we eventually focused on the -20.5
sample and extended the study to DR6.  After the detection of an
\MgII\ absorber system at an impact parameter of 62 h$^{-1}$~kpc, we
extended our impact parameter radius from 65 to 75 h$^{-1}$ kpc.  As
Fig.~\ref{fig:impact_histo} shows and a K-S test suggests, the sample
is deficient in galaxies with impact parameters $\gtrsim 50$ h$^{-1}$
kpc (P$_{\rm K-S} \sim 0.01$).

The targets are listed in Table~1; only galaxy SDSSJ030313.02-001457.4 is
missing from the DR6 database, for reasons that are unclear.  As a
result, we analyze all the spectra in the context of the SDSS VAGC DR6
\Mrh~$\leq -20.5$ sample, referred to as the DR6 -20.5 sample
hereafter.

\subsection{The Observational approach and the data}

We conducted observations on 2008 March 2 and 2009 January 24-25 using
the 1200 l mm$^{-1}$ grism blazed at 3400 \AA\ with LRIS B; the
longslit spectra approximately cover the wavelength range $2910 -
3890$~\AA. The objects for which we use slitmasks cover a similar
area, sometimes beginning just redward of $3000$~\AA.  The grism
has good blue sensitivity, with an efficiency that remains high even
shortward of 3100 \AA.  Its dispersion is 0.24 \AA\ per pixel, which
yields a $\sim$1.6 \AA\ resolution with our 1$\farcs$0 slit and allows
us to resolve the $\lambda$2796, 2803 \AA\ doublet easily.  We took
1-3 exposures for total on-source integration times of 1800~--~3600
seconds, depending on the magnitude of the quasar and the conditions
during the night.  We compute sensitivity limits that reflect these
changing conditions on an object-by-object basis.  We follow standard
data reduction and extraction techniques.  Because neither the sky nor
the quartz lamps provide substantial flux at $\sim$3100~\AA, the data
were not flat fielded.

For the detections, we measure the equivalent widths of 2796~\AA\
(only) by fitting Gaussians directly to the spectra and,
simultaneously, the background.  To estimate the quality of the
spectra and the upper limits for the non-detections on an
object-by-object basis, we compute a running signal-to-noise estimate
of the spectra in the region around the expected \MgII\ absorption at
2796~\AA.  Taking a value that is intermediate between the fit widths
of our first two detections, we assume an 8-pixel ($\sim 1.9~\AA$)
line width for each line and compute the total absorption ``signal''
from within these 8 pixels for each of the the two lines as a function
of potential redshift within 300 km~s$^{-1}$ of the galaxy center.  We
take the continuum from the surrounding area. The assumed noise
estimate includes contributions from the object counts, the
background, readnoise, and an estimate of the error from not
flat-fielding the data.  We assign upper limits based on the
rest-frame equivalent width of a S/N $=3$ detection.  These results
are listed in Table~2.

\begin{figure}
\plotone{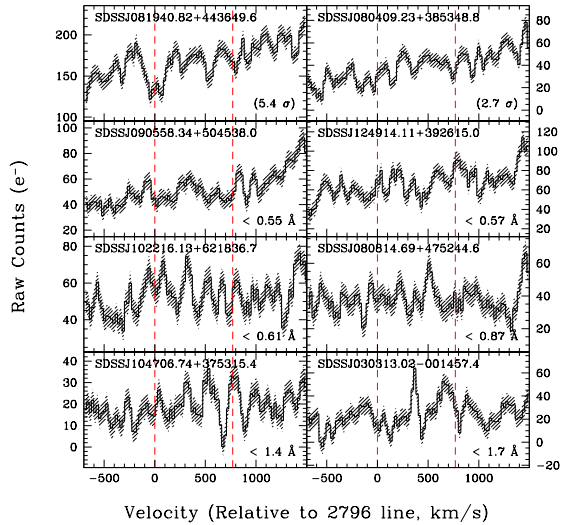}
\caption{The potential detections ({\it Top row}) and weaker
non-detections that are inconclusive with respect to strong
($>0.3$~\AA) absorbers.  We plot raw object counts in detected
electrons as a function of effective rest-frame (peculiar) velocity
shift from the centroid of the 2796 \AA\ line. Shading indicates the
errors and vertical lines ({\it Dashed, red}) indicate the expected
positions of the MgII lines from the galaxies if they occur at the
redshift centroid. We list a short-hand version of the quasar name
({\it Upper left}), and either the significance of the potential
detection or the 3$\sigma$ upper limit of non-detection (rest-frame;
{\it Lower right}). The spectra are box-car smoothed by 3 bins. }
\label{fig:ambiguous}
\end{figure}

\begin{figure}
\plotone{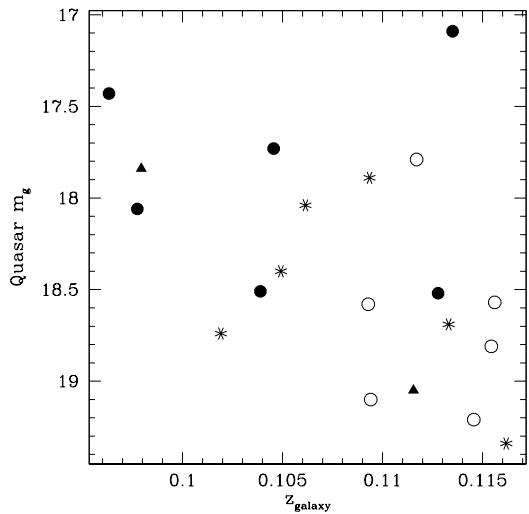}
\caption{Absorber result as a function of galaxy redshift and quasar
$g$-band apparent magnitude.  Using data from
Table~\ref{tab:absorbers}, we plot detections of strong ($> 0.3$~\AA)
absorbers ({\it Filled circles}), non-detections of strong absorbers
({\it Open circles}), potential strong absorbers ({\it Filled
triangles}), and weak non-detections ({\it Stars}) which could be
strong absorbers but not very strong absorbers (see
Table~\ref{tab:absorbers}).  The significant non-detections of strong
absorbers, which require the best data quality, cluster at higher
redshifts because of atmospheric extinction.}
\label{fig:quality}
\end{figure}

Figs.~\ref{fig:detections} - \ref{fig:ambiguous} show the raw-count
spectra and images for the sample classified by whether we find a
significant detection of \MgII\ within 300 km s$^{-1}$ of the systemic
redshift of the galaxy.  We only show images of the clear detections
and non-detections.  Table~2 lists the detection status of the
galaxies, as well.  Two additional sources, SDSSJ080409.23+385348.8
and SDSSJ081940.82+443649.6, are formally detected at the
$\lesssim$3$\sigma$ level in the 2796~\AA\ line but at $\gtrsim
3\sigma$ significance in the combination of both lines, although the
absorbers are not visually evident in the data.  We list these two
sources as ``probable detections.''  Fig.~\ref{fig:quality} describes
the redshifts and quasar magnitudes for the observations, illustrating
that the significant non-detections --- the most stringent
requirements on the spectra --- are much more readily obtained at
relatively high redshifts, as expected from the sharp atmospheric
cutoff near $3050-3100$~\AA.

If we increase the assumed width used to calculate the upper
limits from 8 to 12 pixels, the upper limits increase by $\sim$20\%,
pushing SDSSJ111342.42-000730.7 and SDSSJ140843.77+004730.4 below the
limit.  If a small feature at $\sim$100 \kms\ in
SDSSJ111342.42-000730.7 were real and the blue line were masked by the
unrelated absorption feature at 900 km/s, its \ew\ would still be only
be 0.2~\AA, so it is not a strong absorber. SDSSJ140843.77+004730.4
is a potential (barely) 3$\sigma$ detection, but it would be at $\sim$-316 \kms,
which is probably unphysical.

Most of the galaxies do not have evident faint companions closer to
the quasar.  However, the strong absorber in SDSSJ144033.82$+$044830.9
has a companion with m$_{\rm r}$ = 18.1 and M$_{\rm R} = -19.8$ at a
projected 18.6 h$^{-1}$ kpc from the quasar. Although its redshift was
not acquired as part of the SDSS, we obtained a spectrum with the Keck
telescope on 2009 January 25, measuring $z=0.1128$ after correcting to
the local standard of rest using the IRAF tast RVCORRECT.  This
redshift is within the measurement error of the redshift for the
primary galaxy. The companion is even bluer than the luminous galaxy,
however, with $g-r \sim 0.46$.  Both the companion and the Mg {\sc
II}-absorbing gas are almost certainly within the dark matter halo of
the more luminous galaxy.  Thus, they are all likely associated with
the potential fuel for star formation in the outskirts of the central
galaxy.

\section{Results}

To explore the characteristics of ``strong'' [\ew\ $> 0.3 \AA$] \MgII\
absorbers in the local universe, we compare the properties of the
detected absorbers with the non-absorbers and the full volume-limited
sample.  

\begin{figure}
\plottwo{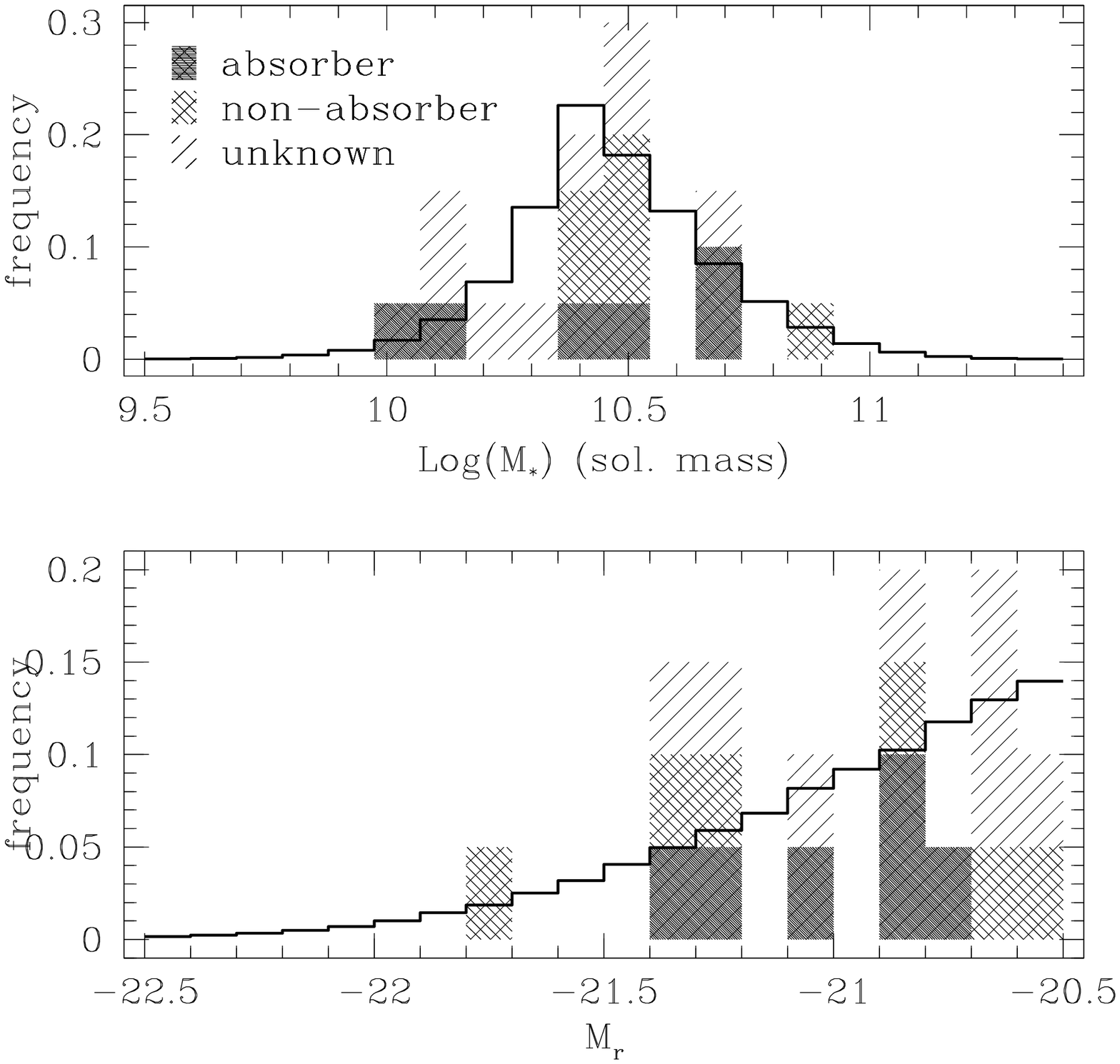}{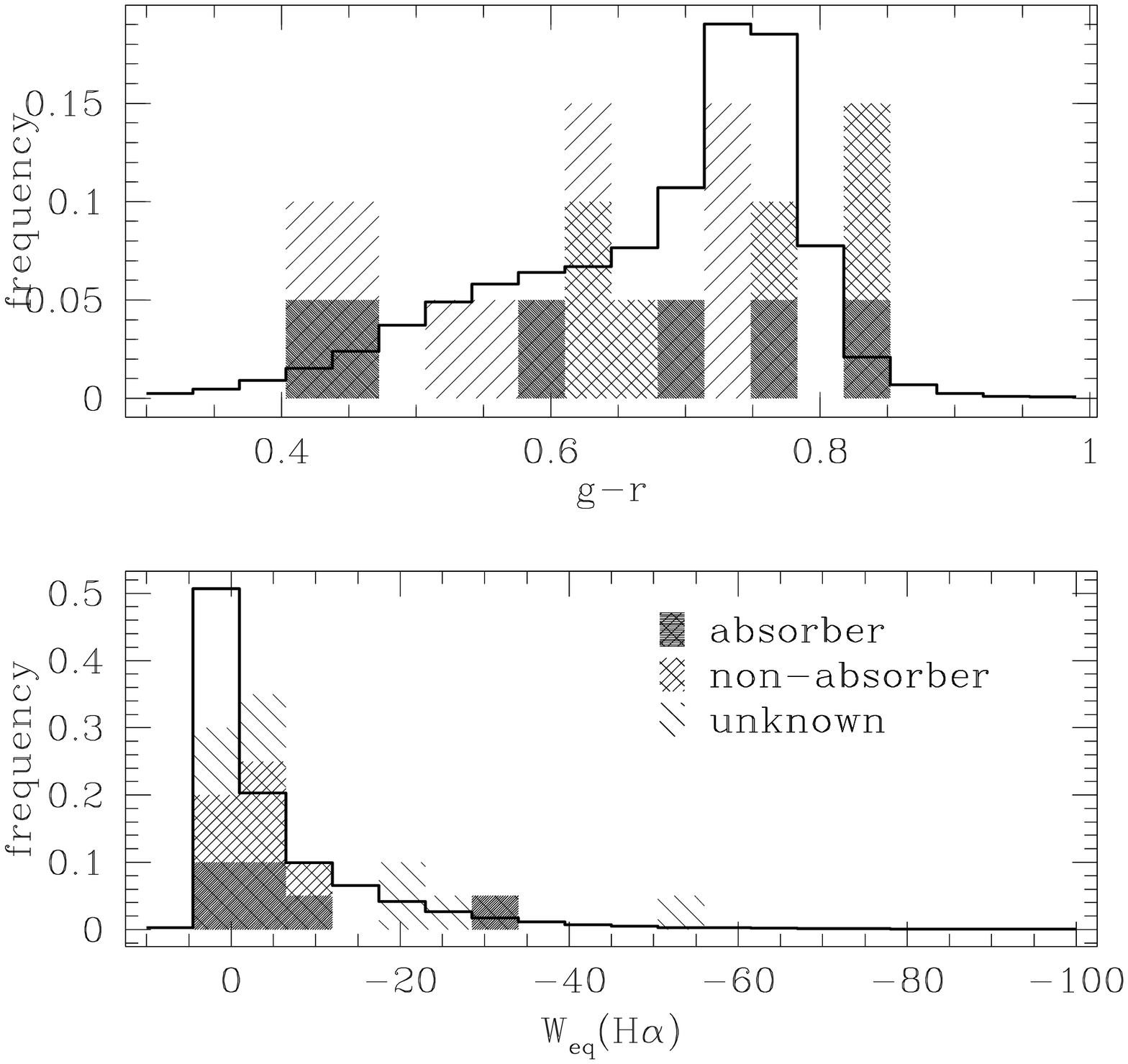}
\caption{Properties of the targeted galaxies in the context of the
volume-limited SDSS sample.  When they are available, we show the
approximate stellar masses ({\it Top left}), $r$-band absolute
magnitudes ({\it Bottom left}), $g-r$ colors ({\it Top right}),
and \ewHa\ ({\it Bottom right}) for the full volume-limited M$_{\rm r}
-5 \log{h} \leq -20.5$ sample ({\it solid line}), the galaxies we have
already targeted.  We show the ambiguous potential detections and weak
non-detections ({\it lightest shading}) the strong non-detections
({\it medium shading}) and the detections({\it darkest shading}).
Here, we plot the
luminous galaxy for the absorber SDSSJ144033.82+044830.9
($\log{M_{\star}} = 10.42, {\rm M_r} = -21.3, g-r = 0.603$), although
its bluer and fainter minor companion is the likely source of at least
some of the \MgII\ absorption. The histograms for the targeted galaxies
are non-overlapping.}
\label{fig:sample}
\end{figure}

Table~1 and Fig.~\ref{fig:sample} show properties of the samples,
including rest-frame colors, luminosities, and stellar masses from the
kcorrect fitting results distributed with the NYU-VAGC
\citep{Blanton05b, Blanton07}.  The solid lines show the full
volume-limited samples, where measurements are available, and the
filled histograms show the probed galaxies. The detected \MgII\
absorbers span the full range of colors and \ewHa, and most of the
range of luminosities and stellar masses of the sample.  For this
small sample of six absorbers, K-S tests reveal no significant
differences between the distributions of the absorbers and the full
sample.  However, we note that K-S tests are not uniformly sensitive to 
all kinds of trends.

K-S tests reveal no evident unfairness in the sampling of parameter
space except in H$\alpha$.  The 20 sampled galaxies fall
preferentially at somewhat smaller \ewHa\ (stronger emission) than the
full volume-limited sample (P$_{\rm K-S} = 0.05$), with an average of
-9.4~\AA\ for the 20 probed systems as compared with -7~\AA\ for the
full volume-limited sample.

\begin{figure}
\plottwo{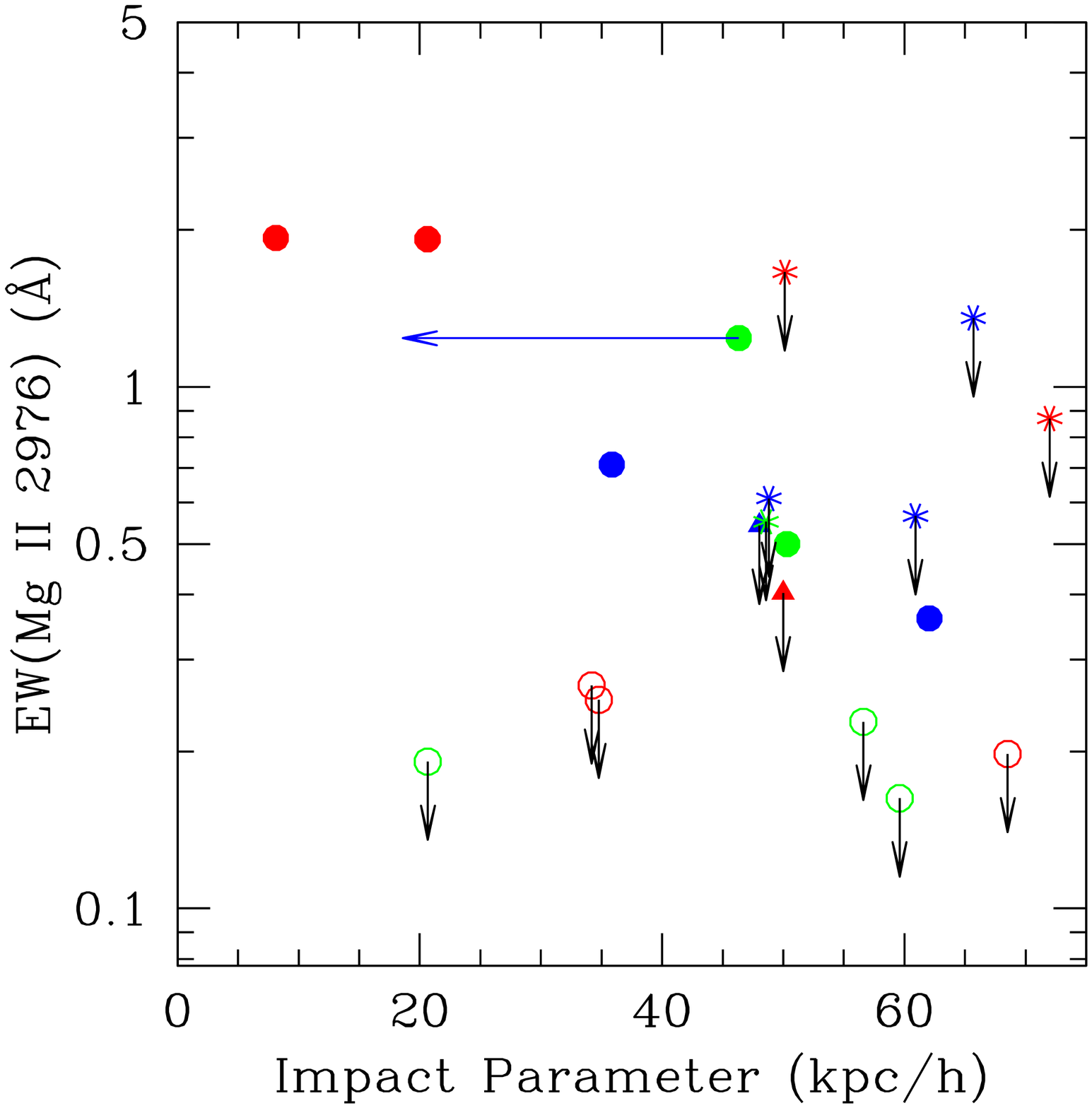}{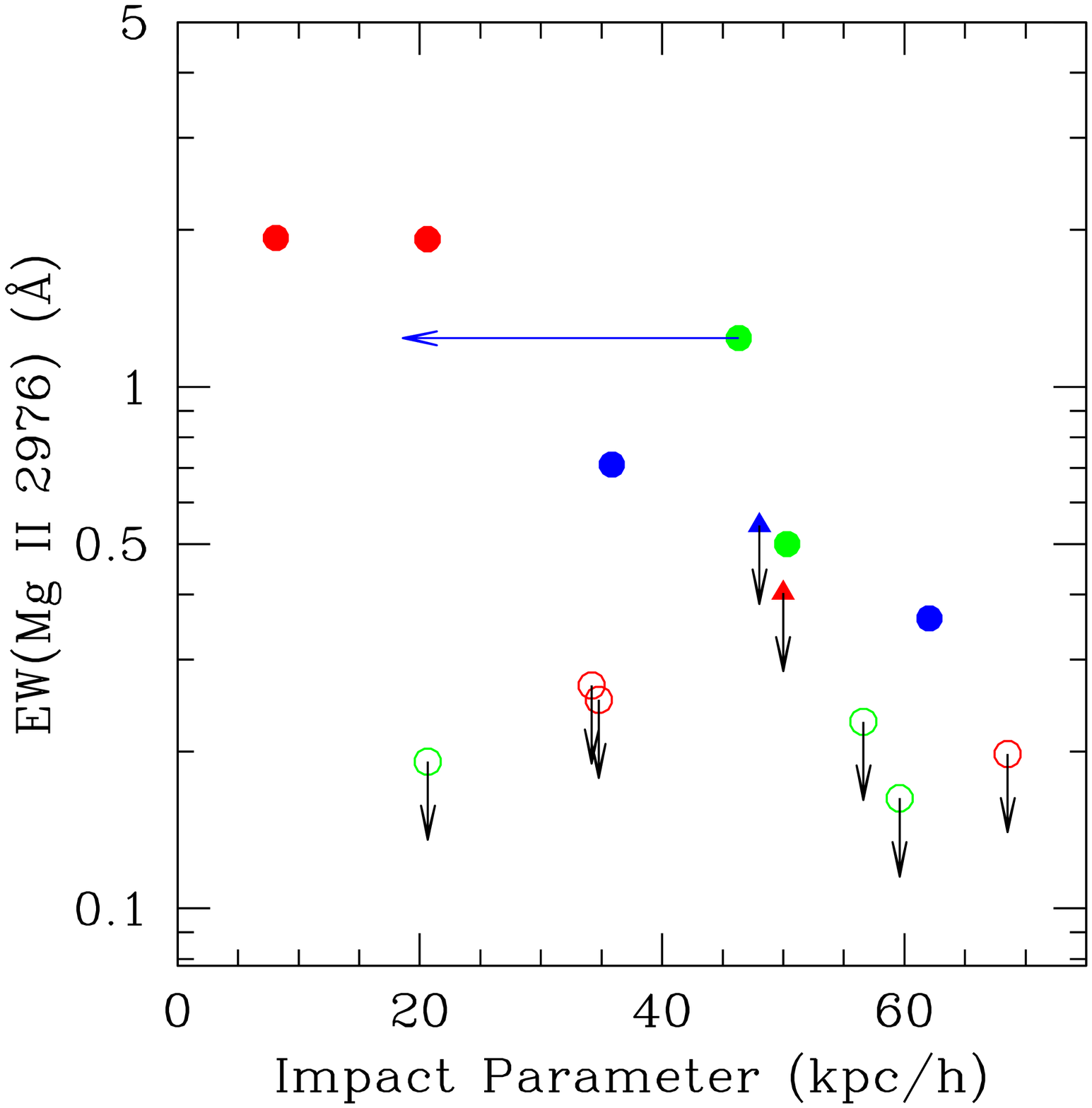}
\caption{ \MgII\ absorption as a function of impact parameter from our
volume-limited SDSS sample.  ({\it Left}) We plot the measured \ew\ or upper limit
of the 2796~\AA\ absorption line as function of impact parameter
between the quasar and the targeted galaxy.  Points are segregated
based on color ({\it Blue: $g-r < 0.6$, green: $0.6 < g-r < 0.7$, 
or red:  $g-r > 0.7$}) and 
on the presence or absence of 2796~\AA\ absorption as described in
Table~\ref{tab:absorbers}.  ({\it Right}) We show the same figure without the
ambiguous, lower S/N spectra.  SDSSJ144033.82+044830.9 is plotted at the luminous
galaxy's position, but with an arrow pointing to the position of its confirmed
minor companion.}
\label{fig:impact}
\end{figure}

\subsection{Star formation}

The data suggest, but do not conclusively establish, the existence of
trends in \MgII\ properties and star-forming properties.  The most
striking potential color dependence on absorption characteristics is
that confirmed non-strong-absorbers ({\it Medium shading}) are green
or red but not blue.  Although the trend does not yield a low K-S
probability, there is a somewhat significant effect in the measured
average $g-r$ color.  The average color for the full sample is $<g-r>
= 0.643$, while it is 0.723 for the non-detections of strong
absorbers.  If we randomly draw six $g-r$ colors at random from the full
volume-limited DR6 sample 10,000 times, the average is only as red as
0.723 by chance 15\% of the time, a $\sim1\sigma$ result.  A larger
sample is crucial for confirming this possible trend.  For \ewHa, the
non-absorbers are relatively weaker star-formers, but at lower
significance.  The average for the full sample is \ewHa\ $= -9.4$~\AA\
while it is -2.7~\AA\ for the 6 non-absorbers, but an average this
large happens by chance 28\% of the time.

Fig.~\ref{fig:impact} depicts the relationships between the colors,
impact parameters, and strengths of the absorbers.  We plot \ew\ as a
function of impact parameter for galaxies with different absorption
status, also color-coding based on $g-r$ color, where blue points have
$g-r < 0.6$, green points have $0.6 < g-r < 0.7$ and red points have
$g-r > 0.7$ . In the right-hand side of the figure, we eliminate the
most ambiguous points to give a clearer view of the possible
trends. 

Among the absorbers, the tendency for stronger absorbers to appear at
smaller impact parameters, as noted by \citet{Steidel95} and more
recently \citet{Zibetti07}, is also suggested by our data.  All of the
bluest galaxies are either strong absorbers or inconclusive, although
the ``green'' galaxies can be non-absorbers.  The confirmed absorbers
at higher impact parameters are all blue.  Furthermore, the only
definitive red absorbers are at very small impact parameter ($\leq
20$~\kpc), although one possible red absorber (SDSSJ081940.82+443649.6)
lies at 48 h$^{-1}$ kpc.  We note that the red absorbers all exhibit
some emission-line flux in their spectra, with [NII]/H$\alpha$ flux
ratios characteristic of LINERs; these red, emitting galaxies are very
common \citep[e.g.,][]{Yan06,Graves07}.

Overall, the data are suggestive that in the low-z universe, the
absorption and covering fraction may depend on the star-forming
properties of galaxies.  Specifically, non-absorbers may be primarily
green or red galaxies, and red absorbers may favor lower impact
parameters.  These relationships are in broad agreement with the
possibility that a gas halo is required to replenish continuously
star-forming galaxies \citep[e.g.,][]{Larson80} and with the
statistical finding of \citet{Zibetti07} that stronger \MgII\
absorbers in SDSS quasars are surrounded by bluer light distributions.
Confirming these results would require better sampling by color that
includes more blue galaxies at lower impact parameters and more red
galaxies at higher impact parameters.

\begin{figure}
\plotone{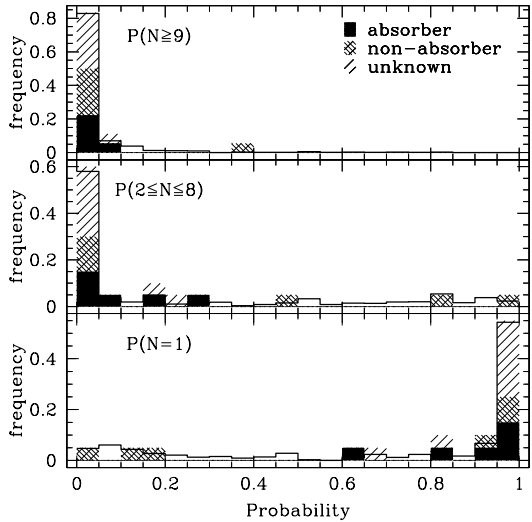}
\caption{The environmental properties of the targeted galaxies in the
context of the volume-limited SDSS sample.  We use numerical
simulations to estimate the probability that the galaxies are ({\it
Bottom}) alone in their dark matter halos with respect to other
luminous galaxies, ${\rm P(N=1)}$, ({\it Middle}) in systems with 2-8
luminous members, ${\rm P(2 \leq N \leq 9)}$, 
or ({\it Top}) in dense systems with $\geq 9$ members
${\rm P(N \geq 9)}$.  The graph types follow Fig.~\ref{fig:sample}.
Here, we plot the results for luminous galaxy near the absorber
SDSSJ144033.82+044830.9, with ${\rm P(N=1)}=0.93$,
${\rm P(2 \leq N \leq 8)}=0.06$, and ${\rm P(N \geq 9)}=0.009$.
}
\label{fig:density}
\end{figure}

\subsection{Environment}

A modification of the method described in \citet{Barton07} allows us
to characterize the environments of the galaxies in the volume-limited
sample objectively, by comparison to a cosmological dark matter
simulation \citep{Zentner03, Zentner05, Allgood06}.  Using
number-density matching to match the \Mrh~$\leq -20.5$ sample to dark
matter halos in the simulation with circular speeds $\geq 240$
km~s$^{-1}$, we create an artificial redshift survey with the
simulation and bin model halos by the distance to the nearest massive
neighbor within 1000 km s$^{-1}$, \DN, and the number of neighbors it
has in total within 700 h$^{-1}$ kpc and 1000 km s$^{-1}$, \Nseven.
Then, for each bin in \DN\ and \Nseven, we compute the probability
that a halo in that bin is alone with respect to other halos in the
sample, P(N=1), the probability that a galaxy in each bin has 1-7
companions, P($2 \leq {\rm N} \leq 8$), and the probability that it
resides in a cluster with $\geq 8$ other halos, P(N$\geq9$).  For each
galaxy in the volume-limited SDSS sample, we measure \DN\ and \Nseven\
from the data then compute P$(N=1)$, P($2 \leq {\rm N} \leq 8$), and
P(N$\geq9$) using the corresponding bin in the simulation.

Fig.~\ref{fig:density} shows the distributions of these probabilities
for the portion of the full volume-limited sample that falls into
relatively complete regions of the SDSS.  Again, because of the small
sample sizes, K-S tests reveal few significant trends; comparing the
distribution of P($2 \leq {\rm N} \leq 8$) shows that the sample of 20
galaxies probed this study as a whole differs at a minor level from
the population at large at (P$_{\rm K-S}=0.10$).

As the distribution of P(N=1) shows, most of the galaxies in the
volume-limited sample are isolated by these criteria, primarily
because the magnitude limit of the sample is quite luminous.
Nonetheless, the detected absorbers are even more isolated than the
population at large, with probabilities ${\rm P(N=1) } \gtrsim$60\%.
We note that although 71\% of the galaxies in the full sample fall
into this regime, the probability that all 6 absorbers fall into the
P(N=1)~$>0.6$ regime by chance is still modest (12\%).  The possible
environmental effects are further accentuated by the results for the
non-absorbers.  Only 18.5\% of the full volume-limited sample have
P(N=1)$< 0.20\%$, while 3/6 of the non-absorbers are in this category.
The binomial probability that 3 or more would fall into the P(N=1)~$ <
0.2$ regime by chance is $\sim8\%$.  Taken together, the two
constitute a $\gtrsim$2$\sigma$ result, although this significance is
somewhat overstated because of the {\it a posteriori} nature of the
estimate.  Thus, the data suggest that absorbers tend to be isolated
and that non-absorbers are more likely to have companions.

This result is broadly consistent with ``strangulation,'' or the
stripping of a galaxy's outer gas halo when it falls into a denser
system \citep[e.g.,][]{Larson80}.  When galaxies become substructure
in a larger system, this stripping eventually cuts off the fuel for
star formation, explaining why galaxies in denser environments
gradually turn red.  As the data suggest, these galaxies in denser
environments may be less likely to exhibit \MgII\ absorption.

\begin{figure}
\plotone{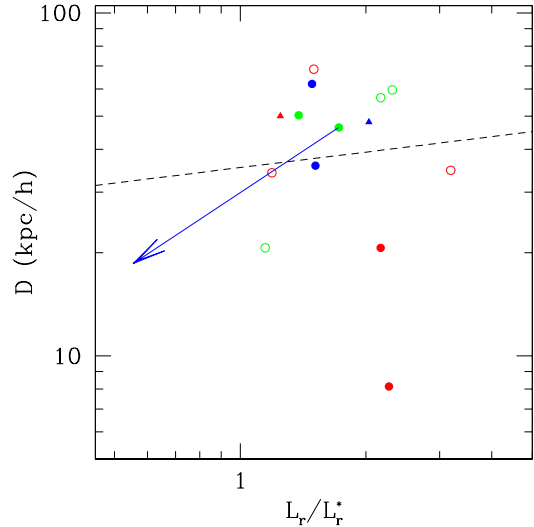}
\caption{ Absorption status as a function of galaxy luminosity and
impact parameter.  We plot the more reliable absorbers ({\it Filled
circles}), non-absorbers ({\it Open circles}), and potential or likely
absorbers ({\it Filled triangles}) as a function of absolute $r$-band
luminosity and impact parameter, following \citet{Steidel95}.  We plot
SDSSJ144033.82+044830.9 at the luminous galaxy's position, but with an
arrow pointing to the position of its confirmed minor companion.
Results differ dramatically from the intermediate-redshift Steidel
study in which galaxies above the line were essentially all
non-absorbers and galaxies below the line were nearly all
absorbers. The points are color-coded as in Fig.~\ref{fig:impact}.}
\label{fig:halo_size}
\end{figure}

\subsection{Covering fraction}

The covering fraction of \MgII\ absorption is a fundamental property
that describes the statistical amount of \MgII-absorbing gas around
typical galaxies; it provides an important constraint on the behavior
of gas in models \citep[e.g.,][]{Kaufmann09,Kacprzak09}.
Hydrodynamic simulations predict a range of \MgII\ covering fractions
depending in detail on the temperature and entropy of the gas.  For
example, Kaufmann et al.\ (2009) show that a change in temperature of
a factor of two can change the \MgII\ covering fraction by a factor of
$\sim$30.  Thus, an accurate and unbiased measure of this fraction
around a uniform set of galaxies, even with accuracy that is within a
factor of a few, is a strong constraint on the nature of halo gas in
galaxies. 

Several empirical studies explore models in which the size of the
absorbing gas ``halo'' scales with galaxy luminosity
\citep{Bowen95,Steidel95,Guillemin97,Kacprzak08}.  Because our
luminosity range is extremely small, we primarily constrain the gas
``halo'' near L$_r^{\star}$ \citep[$\sim$-20.44;][]{Blanton03}. We
show the relationship between absorber status, impact parameter, and
galaxy luminosity in Fig.~\ref{fig:halo_size}.  The dotted line is the
line from \citet{Steidel95}, albeit for $K$-band luminosities. In
their study, almost all of the intermediate-redshift sample are
non-absorbers above the line, while below the line, their galaxies are
almost exclusively absorbers.  Our data are very inconsistent with
this picture, with absorbers and non-absorbers falling above and below
the line in nearly equal numbers.  Furthermore, because one of the
galaxies with the largest impact parameters exhibits \MgII\
absorption, our only constraint on the gas halo radius is that $R
\gtrsim 62$~h$^{-1}$~kpc near L$_r^{\star}$.  This inconsistency could
arise from selection effects, redshift differences, or by chance.

Because of the non-uniformity in the quality of the spectra, we are
forced to cull our sample in order to yield an estimate of f$_{c}$,
the covering fraction of gas or the fraction of sightlines through the
gas ``halo'' that will exhibit ``strong'' [\ew\ $\geq 0.3~\AA$]
absorption \citep{Steidel95}.  To measure the covering fraction of
strong absorbers, we exclude all spectra from Table~2 with 3$\sigma$
upper limits $> 0.3$~\AA, which must necessarily exclude the strong
absorbers SDSSJ092300.67$+$075108.1 and SDSSJ081420.19$+$383408.3, and
could exclude SDSSJ114803.17+565411.4, with an upper limit equivalent
to 0.3~\AA.  These lower-S/N quasar spectra can reveal strong absorbers
but cannot detect strong non-absorbers; thus, their inclusion 
could bias the sample toward finding more absorbers.  We also exclude
SDSSJ144033.82$+$044830.9 because of its minor companion (M$_{\rm R} =
-19.8$) closer to the quasar. Unfortunately, SDSSJ081940.82$+$443649
is an ambiguous case, as the formal 3$\sigma$ limit is $< 0.27$~\AA\
but the detection remains relatively tentative.  Depending on whether
we include or exclude J114803 and/or J144033, we estimate an overall
covering fraction within $\sim$75 h$^{-1}$ kpc of f$_{\rm c}=2/8-4/10
= 0.25-0.4$.  If we consider only those probes within the traditional
$\sim$35 h$^{-1}$ kpc, our statistics become sparse, yielding an
estimate of f$_{\rm c}=0/3-1/4=0-0.25$, depending on how we treat
SDSSJ114803.17+565411.4.

The early studies of \MgII\ absorption suggested that $f_c$ is near
unity.  In our study, several absorbers have larger impact parameters
than some of the non-absorbers.  Our low-$z$ data are inconsistent
with a gas covering fraction of 1. Some previous studies agree with
our results, finding gas covering fractions well below 1
\citep{Bechtold92,Bowen95,Churchill05, Tripp05,Kacprzak08}, while
others favor fractions much closer to 0.8-1 \citep{Steidel95,Chen08}.
Broadly speaking, ``reverse'' studies such as ours, that identify
galaxies first and then explore \MgII\ properties, may measure lower
covering fractions.  The discrepancies among studies could also result
from dramatic differences in redshift, halo mass, and galaxy
properties for the systems that are probed; both factors are likely to
make a difference \citep[e.g.,][]{Bowen95,Tinker08}.

\begin{figure}
\plottwo{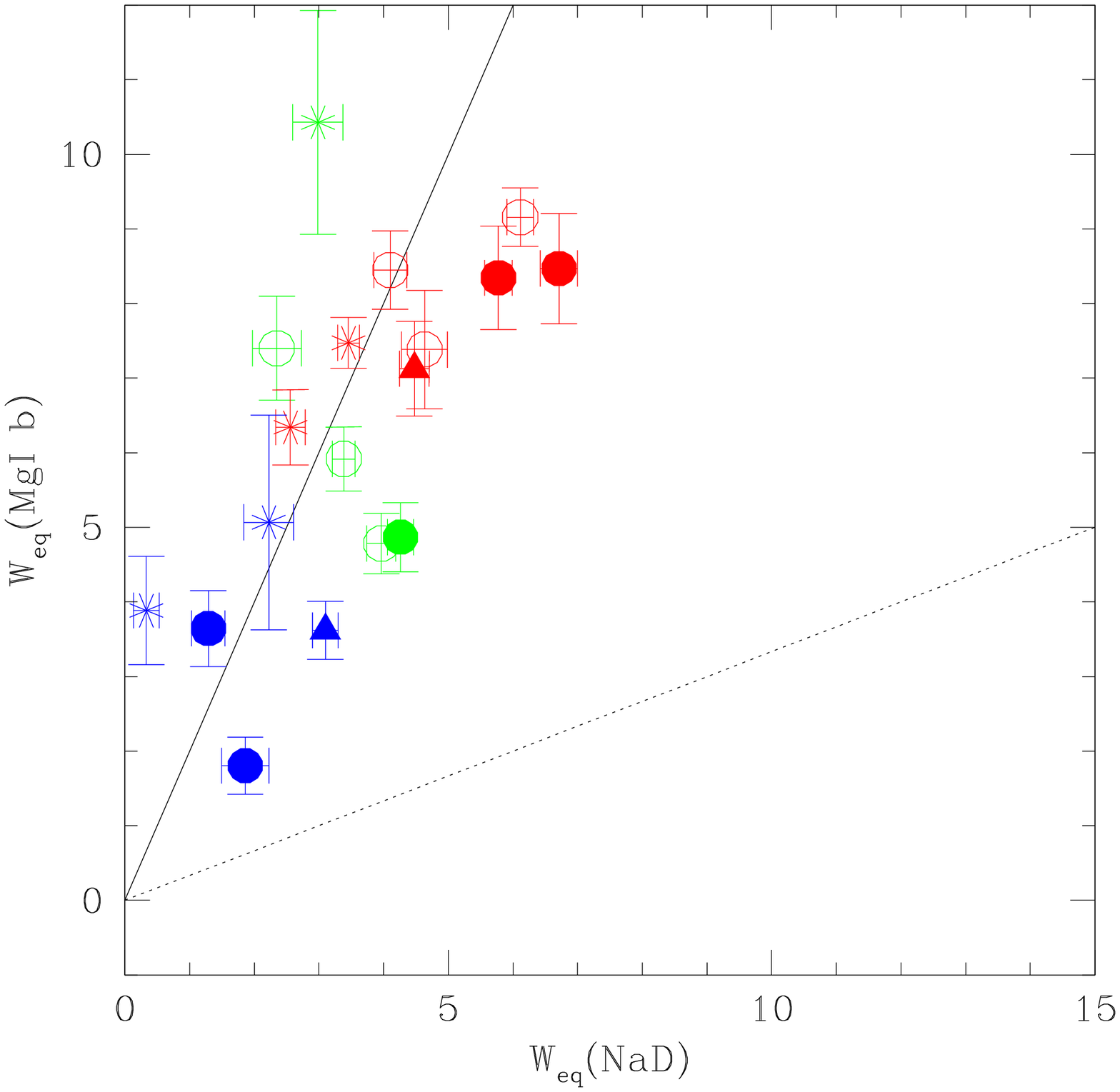}{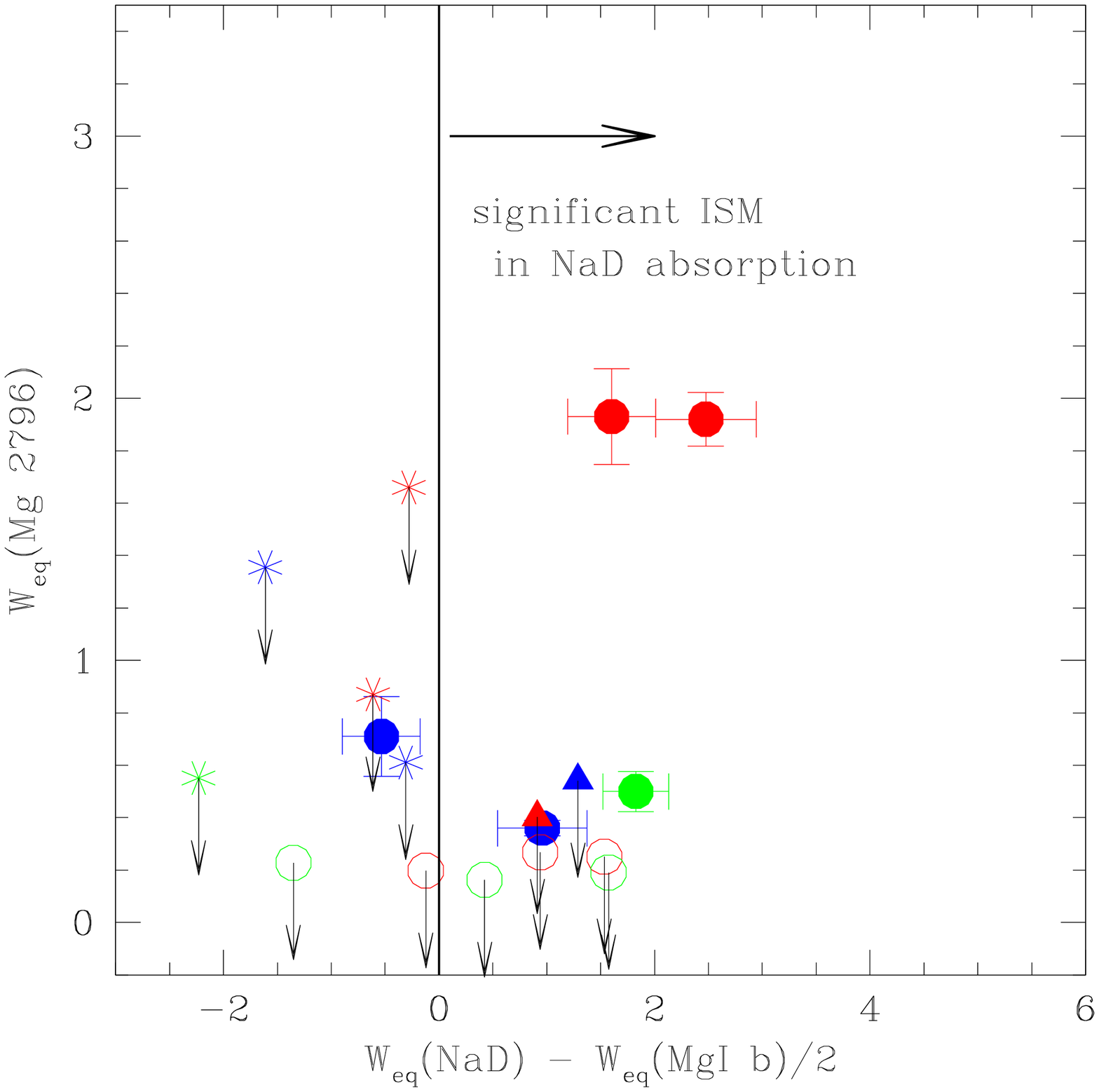}
\caption{The ISM as diagnosed by NaD line strength. Following
\citet{Rupke05a}, we probe the ISM of the targeted galaxies ``down the
barrel,'' using the SDSS spectra of the galaxies themselves.  ({\it
Left}) We plot \ewNa\ vs. \ewMg\ for the galaxies in this study.  The
solid line shows the expected relationship for purely stellar NaD:
\ewNa\ $= 0.5$ \ewMg; the dashed line marks the region where winds are
prevalent \citep{Rupke05a, Rupke05b}.  Thus, galaxies closer to the
right of the solid line are more likely to be ISM dominated in NaD.
({\it Right}) We also plot \ew\ as a function of \ewNa\ corrected for
stellar absorption; the strongest confirmed galaxy halo absorbers
({\it Red points}) have spectra that suggest a significant
contribution of the to NaD line.  We exclude SDSSJ144033.82+044830.9
because of its closer faint companion; point styles follow
Fig~\ref{fig:impact}.}
\label{fig:na}
\end{figure}

\subsection{Other properties of the interstellar medium}

The possible color dependence noted here and elsewhere and the
environmental dependence our data support the expectation that there
are fundamental relationships between the abundance of halo gas in a
galaxy and its star-forming properties. The \MgII\ halo gas may also
relate to the interstellar medium (ISM), as observed
``down-the-barrel'' in galaxy spectra.  The \NaD\ absorption-line
doublet ($\lambda$5890,5896~\AA) is a sensitive, well-studied ISM
diagnostic that falls into the optical regime at low
redshift. Unfortunately, it suffers from contamination by starlight.
Nonetheless, high-resolution studies of the doublet show clear-cut
winds in the \NaD\ line in rapidly star-forming ultraluminous infrared
galaxies \citep{Martin05, Rupke05a, Rupke05b}.

The contamination from a stellar component to the \NaD\ line can be
estimated from the equivalent width of the \MgIb\ triplet
($\lambda$5167,5173,5185~\AA).  Various authors assume a stellar
contribution to \ewNa\ of $(\frac{1}{3} - \frac{1}{2}) \times$ \ewMg.
In Fig.~\ref{fig:na}, we plot \ewMg\ as a function of \ewNa.
Following \citet{Rupke05a}, we plot the line \ewNa\ $=0.5$ \ewMg\
({\it Solid line}); points further to the right of the solid line are
more likely to have contributions from the ISM to the \NaD\
absorption. The dotted line is the approximate location of
starbursting galaxies with strong winds observed in the \NaD\ feature
for the \citet{Rupke05a} sample.  We code the points based on \MgII\
absorption status in the halo of the galaxy.  None of the DR6 -20.5
sample galaxies observed thus far lie in the strong-wind
regime. Nonetheless, there is a suggestion that the galaxies with
\MgII\ halo absorption lie at relatively large \ewNa-to-\ewMg\ ratio,
as illustrated by the right-hand side of the plot.  All of the red
absorbers have relatively strong \NaD\ absorption given their \ewMg.
In fact, only one absorber lies at \ewMg $> 0.5$\ewNa and it is blue
(SDSSJ081420.19+383408.3).  Perhaps red galaxies with halo \MgII\ must
actually be either very dusty or gas-rich enough to exhibit
significant \NaD\ from the ISM which has not yet resumed star
formation.  Blue galaxies with little interstellar \NaD\ could be
systems at the end of their star-forming cycle that have used up or
cleared out their low impact parameter \NaD\ gas, but have retained
their outer halo gas; the relatively strong Balmer absorption in
SDSSJ081420.19+383408.3 [\ewHd\ $=3.6$~\AA] supports this hypothesis.

\section{Conclusions} \label{sec:conclusion}

We use ground-based spectroscopy with LRIS-B on the Keck I telescope
to evaluate the \MgII\ absorption characteristics of a sample of
$\sim$L$^{\star}$ galaxies selected as cleanly as possible from a
volume-limited subsample to M$_{\rm r} + 5 \log{h} \gtrsim -20.5$ of
the SDSS DR6 Value Added Galaxy Catalog \citep{Blanton03}.  We target
20 systems, achieving a range in the quality of spectral observations
that reveals 6 ``strong'' [\ew\ $> 0.3 $~\AA] absorbers and 6
non-absorbers.  We find the following:

\begin{enumerate}

\item Ground-based observations of \MgII\ absorption are possible 
--- with relatively low S/N ratio --- to redshifts as low as $z \sim 0.1$
with modest amounts of observing time.

\item The overall covering fraction, $f_c$, of gas capable of creating
strong ($\geq 0.3$~\AA) \MgII\ absorption must be $< 1$ at $z \sim
0.1$, broadly consistent with some previous studies.  Even inside the
traditional absorption halo radius of $\sim$35 h$^{-1}$ kpc, the
covering fraction of strong absorption is $< 1$, with a naive estimate
of f$_{\rm c} \lesssim 0.25$.

\item The \MgII\ detection in the quasar with the largest impact
parameter lies 62 h$^{-1}$ kpc from the center of the associated
galaxy.  Thus, at least in rapidly star-forming $z \sim 0.1$ galaxies,
strong, but possibly unfilled, \MgII\ absorption gas halos can extend to
at least this distance from a luminous galaxy.

\item The data suggest, at the $\lesssim$2$\sigma$ significance
level, that the \MgII\ absorption properties of the outer halos of
galaxies may depend on the larger-scale environment of the absorbing
galaxy. Absorbers appear to favor low-density environments, while
non-absorbers show a possible preference for denser regions.  This
result is consistent with ``strangulation'' models in which satellite
galaxies are red because their outer gas halos are stripped when they
become substructure.

\item The relationships between galaxy color, impact parameter, and
absorbing halo \ew\ are consistent with a model in which red galaxies
with \MgII\ absorption are typically LINERs which are observed at smaller
impact parameters.  Blue galaxies are possible \MgII\ absorbers to
very large impact parameters (at least 62 h$^{-1}$ kpc); in fact,
all of the bluest galaxies in this study are either strong absorbers or
inconclusive.  These color trends are consistent with the
results from statistical image-stacking studies \citep{Zibetti07},
and with the broad expectations of theoretical models.

\end{enumerate}

\acknowledgments We gratefully acknowledge the Sloan Digital Sky
Survey and the hard work of M. Blanton and collaborators to create the
NYU-Value Added Galaxy Catalog.  We thank James Bullock, Andrew
Zentner, Risa Wechsler, David Tytler, Aaron Barth, Glenn Kacprzak,
and Taotao Fang for assistance, advice and
support.  EJB and JC acknowledge generous support from the Center for
Cosmology at UC Irvine and JC gratefully acknowledges support from
Gary McCue.  The authors wish to recognize and acknowledge the very
significant cultural role and reverence that the summit of Mauna Kea
has always had within the indigenous Hawaiian community.  We are most
fortunate to have the opportunity to conduct observations from this
mountain.

Funding for the Sloan Digital Sky Survey (SDSS) and SDSS-II has been
provided by the Alfred P. Sloan Foundation, the Participating
Institutions, the National Science Foundation, the U.S. Department of
Energy, the National Aeronautics and Space Administration, the
Japanese Monbukagakusho, and the Max Planck Society, and the Higher
Education Funding Council for England. The SDSS Web site is
http://www.sdss.org/.  The SDSS is managed by the Astrophysical
Research Consortium (ARC) for the Participating Institutions. The
Participating Institutions are the American Museum of Natural History,
Astrophysical Institute Potsdam, University of Basel, University of
Cambridge, Case Western Reserve University, The University of Chicago,
Drexel University, Fermilab, the Institute for Advanced Study, the
Japan Participation Group, The Johns Hopkins University, the Joint
Institute for Nuclear Astrophysics, the Kavli Institute for Particle
Astrophysics and Cosmology, the Korean Scientist Group, the Chinese
Academy of Sciences (LAMOST), Los Alamos National Laboratory, the
Max-Planck-Institute for Astronomy (MPIA), the Max-Planck-Institute
for Astrophysics (MPA), New Mexico State University, Ohio State
University, University of Pittsburgh, University of Portsmouth,
Princeton University, the United States Naval Observatory, and the
University of Washington.

\bibliography{Barton09.bbl}

\end{document}